\documentclass[12pt]{article}
\pdfoutput=1
\usepackage{youngtab}
 \usepackage[nottoc]{tocbibind}
\usepackage[nosort]{cite}
\usepackage[colorlinks=true,linkcolor=blue,urlcolor=magenta, citecolor=blue]{hyperref}
\usepackage{fullpage}
\usepackage{amssymb,graphicx}
\usepackage{amsmath}
\usepackage[margin=0.5cm]{caption}
\usepackage{hyperref} 
\usepackage{cite}
\usepackage{upgreek}

\def\e{{\rm e}}

\def\d{\partial}
\def\l{\left(}
\def\r{\right)}

\newcommand{\be}{\begin{equation}}
\newcommand{\ee}{\end{equation}}
\newcommand{\bea}{\begin{eqnarray}}
\newcommand{\eea}{\end{eqnarray}}
\newcommand{\bg}{\begin{gather}}
\newcommand{\eg}{\end{gather}}
\newcommand{\bseq}{\begin{subequations}}
\newcommand{\eseq}{\end{subequations}}

\begin{document}
\begin{titlepage}
\begin{center}
{\Large\bf Yang--Mills Glueballs as Closed Bosonic Strings
%\footnote{Or {\it ``Physical Implications of a 750 MeV Boson" }}
}\\
\vspace{.2cm}
\vspace{0.8cm}
{ \large
Sergei Dubovsky$^{a,b}$ and Guzm\'an Hern\'andez-Chifflet$^{a,c}$ 
}\\
\vspace{.45cm}
{\small  \textit{   $^a$Center for Cosmology and Particle Physics,\\ Department of Physics,
      New York University\\
      New York, NY, 10003, USA}}\\ 
\vspace{.25cm}      
{\small  \textit{$^b$Perimeter Institute for Theoretical Physics, \\Waterloo, Ontario N2L 2Y5, Canada}}\\ 
\vspace{.25cm}      
      {\small  \textit{   $^c$Instituto de F\'isica, Facultad de Ingenier\'ia,\\ Universidad de la Rep\'ublica,\\
      Montevideo, 11300, Uruguay}}\\ 
\end{center}
\begin{center}
\begin{abstract}
We put forward the Axionic String Ansatz (ASA), which provides a unified description for the worldsheet dynamics of confining strings in pure Yang--Mills theory both in $D=3$ and $D=4$ space-time dimensions.
The ASA is motivated by the excitation spectrum of long confining strings, as measured on a lattice, and by recently constructed integrable axionic  non-critical string models. 
According to the ASA, pure gluodynamics in 3D is described
by a non-critical bosonic string theory without any extra local worldsheet degrees of freedom.
We argue that this assumption fixes the set of quantum numbers (spins, $P$- and $C$-parities) of almost all glueball  states.
We confront the resulting predictions with the properties of approximately $1^2+2^2+3^2+5^2=39$ lightest glueball states measured on a lattice and find a good agreement. On the other hand, the spectrum of low lying glueballs in 4D gluodynamics suggests the presence of a massive pseudoscalar mode on the string worldsheet, in agreement with the ASA and lattice data for long strings.

%Properties of confining flux tubes, as observed on a lattice, suggest that pure gluodynamics in 3D may be described
%by a non-critical bosonic string theory without any extra massless or massive worldsheet degrees of freedom.
%We argue that this assumption fixes the set of quantum numbers (spins, $P$- and $C$-parities) of all glueball  states.
%We confront the resulting predictions with the properties of approximately $1^2+2^2+3^2+5^2=39$ lightest glueball states measured on a lattice and find an excellent agreement. On the other hand, the spectrum of low lying glueballs in 4D gluodynamics suggests the presence of a massive pseudoscalar mode on the string worldsheet, confirming previous long string results.

\end{abstract}
\end{center}
\end{titlepage}
%\end{document}
\tableofcontents
\newpage
\section{Introduction}
Solving the $SU(N_c)$ Yang--Mills theory in the planar limit is an old, fascinating and famously hard puzzle. Despite an impressive progress achieved in understanding of certain aspects of this problem, such as the discovery of AdS/CFT \cite{Maldacena:1997re,Gubser:1998bc,Witten:1998qj} and of integrability in ${\cal N}=4$ supersymmetric Yang--Mills \cite{Beisert:2010jr}, a solution of confining non-supersymmetric planar gluodynamics remains elusive. One of the goals of this paper is to convince the reader that the subject still presents a fundamental interest, and that now may be a good time to revisit it. We will also report on a partial progress, mostly for the 3D version of the problem.

The major reason for the interest in this problem is the known project of the answer, which goes back to early 70's \cite{'tHooft:1973jz}.
Namely, it is widely expected that gluodynamics allows a dual description in terms of a string theory, which becomes weakly coupled at large number of colors $N_c$. So to large extent  the challenge ``reduces" to constructing a worldsheet theory of confining strings in the free ($N_c=\infty$) limit. 

It is important to keep in mind that despite numerous strong arguments favoring the possibility of a weakly coupled string description for large $N_c$ gluodynamics, this conclusion is far from being granted (see, e.g., \cite{Polchinski:1992vg,Cordes:1994fc,Polyakov:1997tj}, for nice sets of arguments going both ways). In particular, it was advocated \cite{Polchinski:1991tw} that even if the string reformulation exists, it is likely to be quite complicated and to  involve an infinite set of worldsheet fields. As we argue here, this conclusion may be too pessimistic, and there is evidence suggesting that the structure of the worldsheet theory is quite simple.

To make the task of solving planar gluodynamics more tractable and  better defined we divide the question into two parts.
This separation allows to sharpen the notion of worldsheet theory of confining strings, and makes it manifest that such a theory exists. 

As a first step, following the discussion in \cite{Dubovsky:2015zey}, we consider a sector of the theory with an infinitely long fluxtube stretching through the whole of space. Heuristically, such a sector exists, because in the absence of fundamental quarks a fluxtube cannot break. More formally, a pure glue theory is invariant under a global 1-form symmetry \cite{Gaiotto:2014kfa} (the center symmetry \cite{'tHooft:1977hy}), 
and an infinite flux tube (in the fundamental representation) carries a unit charge with respect to this symmetry. To avoid involving  
 infinite energy states into the discussion, one may introduce an IR regulator by compactifying one of spatial dimensions on a circle, which is considered much larger than any other distance in the problem. Then a ``long" flux tube state is created by a Polyakov loop \cite{Polyakov:1978vu} wrapping around the circle\footnote{Note that in the present context the circle corresponds to a compact spatial direction, rather than to the  thermal Euclidean circle as in \cite{Polyakov:1978vu,Susskind:1979up}.}.
 
 Low energy dynamics in this sector is described by a two-dimensional effective theory of modes localized on the flux tube. These necessarily include gapless Goldstone modes arising as a result of spontaneous breaking of the bulk Poincar\'e symmetry $ISO(1,D-1)$ to the product of the  wordsheet Poincar\'e and of the transverse rotations $ISO(1,1)\times O(D-2)$ (here $D$ is the total number of space-time dimensions).
At small $N_c$ the UV cutoff $\Lambda_{UV}$ of this low energy theory is set by the mass $m_g\sim\Lambda_{QCD}$ of the lightest glueball. At energies above $m_g$ two-dimensional unitarity is violated by the production of bulk glueball states.

Things get more interesting at large $N_c$ because the production of bulk glueballs is now suppressed and as a result the UV cutoff $\Lambda_{UV}$ is parametrically higher than $m_g$. Let us stress that by $\Lambda_{UV}$ we understand the scale where a violation of
  two-dimensional unitarity is a real physical effect, rather than an artifact of a perturbative expansion. For any $N_c$  perturbative two-dimensional unitarity is broken at energies around $\Lambda_{QCD}$, so that the two-dimensional theory is necessarily strongly coupled in the energy range $\Lambda_{QCD}<E<\Lambda_{UV}$.

\begin{figure}[t!!]
	\centering
	\includegraphics[width=0.5\textwidth]{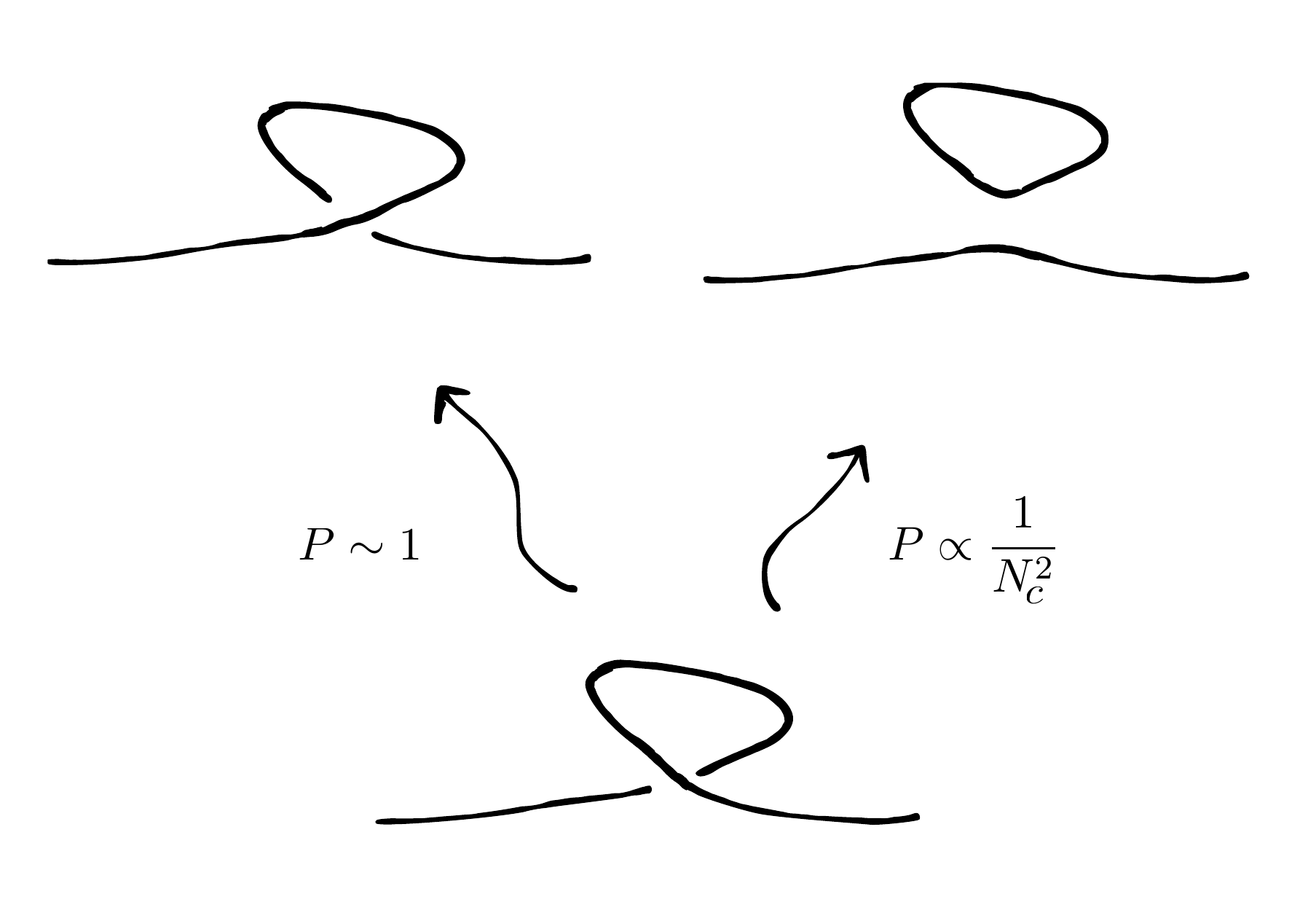}
	\caption{Glueball production in the worldsheet scattering proceeds through string intersection and interconnection. At large $N_c$ the probability of interconnection is suppressed by $1/N_c^2$.}
		\label{fig:intersection}
\end{figure}

Semiclassically bulk glueball production  proceeds through string self-intersections followed by interconnections and by pinching off of a closed string loop (see Fig.~\ref{fig:intersection}).
In four dimensions the cutoff  energy $\Lambda_{UV}$ was estimated  \cite{Dubovsky:2015zey} to
scale as
\be
\label{Luv}
\Lambda_{UV}\sim {N_c\over \ell_s}\;,
\ee
where $\ell_s\sim m_g^{-1}$ is the string width, which is related to the Regge slope $\alpha'$ and to the dynamical QCD scale $\Lambda_{QCD}$ as
\[
\alpha'\equiv {\ell_s^2\over 2\pi}\sim \Lambda_{QCD}^{-2}\;.
\]
It is straightforward to adopt the estimate of  \cite{Dubovsky:2015zey} to the 3D case with the same result (\ref{Luv}) for the cutoff.
The main difference between the 3D and the 4D cases is that in 3D a string self-intersection, once formed, stays for a macroscopically long time in the absence of interconnections. This affects the mass spectrum of produced glueballs without changing the cutoff scale.

The principal conclusion from the above discussion is  that the $N_c\to \infty$ limit at fixed energy results in a two-dimensional theory, which gets strongly coupled at $E\sim\ell_s^{-1}$, but stays unitary up to arbitrary high energies. It is this theory that we call the worldsheet theory of confining strings.
The first step in understanding  planar gluodynamics amounts to solving (to a reasonable extent) the worldsheet theory.

Then the second step should be to reconstruct the glueball spectrum from a known worldsheet theory. It is important to stress that, as posed here, the first question should necessarily have an answer. Indeed, with a powerful enough computer one may measure any physical observable 
in the worldsheet theory with an arbitrary accuracy by performing a usual Monte--Carlo simulation of $SU(N_c)$ Yang--Mills theories with larger and larger $N_c$.
On the other hand, it might happen that even the full knowledge of the worldsheet theory is not enough to reconstruct the  spectrum of glueballs, so that the second step cannot be implemented. A trivial example of how this might happen would be if the bulk theory contained an additional sector, which decouples from gauge interactions at large $N_c$. Then the corresponding states cannot be described as string excitations. One of our goals here is to provide  evidence that this does not happen in 3D Yang--Mills theory
and that {\it all} of the glueball states correspond to string excitations.

Of course, to move beyond the strict planar limit the third step in the program should be to calculate perturbative string amplitudes
between glueball states. This goes beyond the scope of the present discussion.

As formulated, the two questions above still sound extremely challenging to solve. We feel that the most realistic way to make progress at the current stage is to follow the experimental guidance. Recall, that even in the simplest case of  2D Yang--Mills, where the string description is to a large extent known \cite{Gross:1992tu,Gross:1993hu,Cordes:1994fc}, it was obtained by making  heavy use of experimental data (the exact ``non-stringy" solution \cite{Migdal:1975zg,Rusakov:1990rs,Witten:1991we} in that case)\footnote{Note, however, that the answers to our two questions in 2D are trivial---the theory exhibits neither worldsheet excitations nor glueballs. The non-trivial (and tractable in 2D) part of the problem is the third question---developing a perturbative string expansion at finite $N_c$.}. 

Unfortunately, in higher dimensions we do not have a luxury of knowing the exact solution, so the only available data to guide us is coming from lattice simulations. Luckily, a significant amount of high quality lattice data was collected recently on the properties of long strings in 3D and 4D large $N_c$ Yang--Mills \cite{Teper:2009uf,Athenodorou:2010cs,Athenodorou:2011rx,Athenodorou:2013ioa,Athenodorou:2016kpd}, as well as on the glueball spectra (especially in 3D) \cite{Meyer:2003wx,Meyer:2004gx,Lucini:2010nv,Athenodorou:2016ebg}, see \cite{Lucini:2012gg} for a review.
As we review in section~\ref{sec:long}, combining these data with modern theoretical tools (such as the Thermodynamic Bethe Ansatz (TBA)\cite{Zamolodchikov:1989cf,Dorey:1996re})  allows to extract a considerable amount of information about the worldsheet theory both in 3D and 4D \cite{Dubovsky:2013gi,Dubovsky:2014fma,Dubovsky:2015zey}.

At this stage it would be definitely  premature to claim the full knowledge of the worldsheet theory either in 3D or in 4D. However, there is already enough information to formulate a non-trivial unified strawman Ansatz for its general structure \cite{Dubovsky:2015zey}, which is consistent with the current long string data.
For reasons explained later, we refer to the resulting theories as axionic strings.
 In particular, in 3D the Axionic String Ansatz (ASA) says that the worldsheet theory does not carry any new local degrees of freedom in addition to the gapless translational mode of the string. 
 % (plus a certain integrable asymptotics in the UV, see section~\ref{sec:long}). 
Given the wealth of data on the spectrum of glueballs (``short" strings), it would be extremely helpful if we could translate this conjecture into predictions about the glueball spectrum. This is precisely the second question above, and the main goal of the present paper is to start addressing this challenge. 

In the present paper we will make a first step in this direction. We will not attempt to calculate the glueball masses, but rather restrict to  
 their possible quantum numbers, which are spins and $P$- and $C$-parities. For the resulting predictions to be useful we will need to make a certain general assumption about the overall structure of the mass spectrum, which will be inspired by what we know about critical strings (or, alternatively, by assuming that the theory is in the proximity of a certain integrable theory). As we explain in section~\ref{sec:spectrum} the ASA allows then to fix quantum numbers of almost all glueballs. This should not be too surprising.
For instance, in the long string sector the assumption that no additional fields are present on the worldsheet immediately fixes the structure of the Kaluza--Klein spectrum, when the string is compactified on a circle. Here we generalize this observation 
to the case of rotating strings.
This result may be considered as a finite volume version of the Goldstone theorem. 

For $39=1^2+2^2+3^2+5^2$ lightest glueball states we compare the resulting predictions with  the most recent lattice data \cite{Athenodorou:2016ebg} in section~\ref{sec:lattice}.
 To our opinion, we find a very encouraging agreement\footnote{It should be noted though that, as a consequence of the rotation symmetry,
some of the states necessarily come in degenerate pairs. Hence the actual number of independent non-trivial predictions to test is smaller than 39.}. Apart from a few of the heaviest states, which are missing in the current lattice data, there is a single pair of states which disagrees with our predictions. Namely, one of the doublets identified as
$3^+$ in \cite{Athenodorou:2016ebg} is predicted to be $1^+$ (see section \ref{sec:spectrum} for a detailed explanation of how the states are labeled). This kind of misidentification is not completely implausible, given that the lattice breaks the continuous rotation symmetry to a discrete subgroup, so spin determination is a very delicate task. At any rate, given that currently this is the only disagreement, falsifying this prediction
will be very interesting, independently of the outcome. We also present the ASA prediction for the quantum numbers of the next $8^2$ states, out of which very few have been measured so far.
Finally, we take a brief look at the 4D glueball spectrum, and find that it strongly suggests the presence of an additional massive pseudoscalar worldsheet state in agreement with the long string data and with the ASA.

In section~\ref{sec:Higgs} we compare the character describing the bosonic string spectrum described here to the linear dilaton character, and to the character of critical bosonic strings analytically continued into $D=3$. In section~\ref{sec:anyons} we discuss several general puzzles concerning the properties of bosonic strings at $D=3$ and comment on a possible resolution. We present our conclusions and discuss future directions in section~\ref{sec:last}.
\section{Long Flux Tubes and the ASA}
\label{sec:long}
Let us start with a lightning review of what is currently known about long confining flux tubes.
For details we refer the reader to \cite{Dubovsky:2013gi,Dubovsky:2014fma,Dubovsky:2015zey}. 

From the bottom up perspective long strings are the natural starting point to study string dynamics, even though this is not how fundamental string theory has been developed. Indeed, both long and short strings can be thought of as systems of Goldstone bosons emerging as a result of spontaneous breaking of the target space Poincar\'e symmetry. However, long strings provide a more conventional quantum field theory setting by preserving  translational invariance along the worldsheet. 

Lattice simulations provide information about the finite volume spectrum of this theory on a circle of size $R$. To extract  information about the worldsheet theory these results have to be compared with theoretical calculations of the finite volume spectrum in the low energy effective theory describing dynamics of long wavelength perturbations on the string worldsheet. Traditionally these calculations have been performed using the $\ell_s/R$ expansion\cite{Luscher:1980ac,Luscher:2004ib,Aharony:2010db,Aharony:2011ga,Dubovsky:2012sh,Aharony:2013ipa}. As a consequence of the non-linearly realized Poincar\'e  symmetry this expansion exhibits a remarkable degree of universality---for instance, in the absence of extra gapless fields on the worldsheet, the first non-universal contribution to the ground state energy of the long string arises at ${\cal O}\l{\ell_s^6/ R^7}\r$ order. This universality is a very unwelcome feature for our purposes---at first sight it makes it  very hard  
to use lattice data as a probe of the worldsheet theory. For instance, it is very challenging to extract any information from the ground state data apart from the absence of extra massless (or anomalously light) fields on the worldsheet.

However, the situation is different for excited states. The universality  of the $\ell_s/R$ expansion still holds of course,
however the radius of convergence of this asymptotic  expansion turns out to be too small here. As a result
 the $\ell_s/R$ expansion  is practically useless for interpreting the available lattice results for excited states. This complication stimulated the development of an alternative perturbative approach. The idea is to calculate first the worldsheet $S$-matrix, and then to relate it to the finite volume spectrum using the TBA technique. 
 
 The TBA method drastically improves the convergence and leads to the identification of a massive pseudoscalar excitation (the worldsheet axion) on the worldsheet of confining strings in 4D. The leading order pseudoscalar interaction of the worldsheet axion with translational modes 
takes the form
\be
\label{S1a}
S_a={Q_a\over 4}\int d^2\sigma \sqrt{-h}h^{\alpha\beta}\epsilon_{\mu\nu\lambda\rho}\d_\alpha t^{\mu\nu}\d_\beta t^{\lambda\rho}a
\ee
where $a$ is the axion field, $X^\mu$ are the embedding coordinates of the string worldsheet,
$h_{\alpha\beta}$ is the induced metric on the worldsheet and
\[
t^{\mu\nu}={\epsilon^{\alpha\beta}\over\sqrt{-h}}\d_\alpha X^\mu\d_\beta X^\nu\;.
\]
Neglecting non-perturbative effects on the worldsheet, the interaction (\ref{S1a}) does not break the axion shift symmetry, because $a$ is coupled to the topological density, which calculates a self-intersection number of the string worldsheet \cite{Polyakov:1986cs}. For the $SU(3)$ gauge group the axion mass was measured to be
\[
\mu_a\approx1.85^{+0.02}_{-0.03}\ell_s^{-1}
\]
and the coupling constant is
\be
\label{Ql}
Q_a\approx 0.38\pm 0.04\;.
\ee

Instead, the 3D lattice data does not require any massive excitations on the worldsheet, although it does show the presence of non-universal corrections to the scattering phase shift.

At first sight this information is hardly enough to come up with a sensible Ansatz for the structure of the worldsheet theory, especially in 3D.  However, the data might be more telling than it appears. To see this, we need to review some further theoretical background.

Motivated in part by the discovery of integrability in ${\cal N}=4$ supersymmetric Yang--Mills, the following question was raised in \cite{Cooper:2014noa,Dubovsky:2015zey}. Is it possible that the worldsheet dynamics in a confining gauge theory is integrable, {\it i.e.} all scattering amplitudes on the worldsheet are purely elastic? 

A partial answer to this question is that integrability necessarily requires the presence of additional massless degrees of freedom on the worldsheet at $D\neq 3, 26$.
Indeed, the Nambu--Goto theory is integrable at the classical level. However the integrability is broken by a quantum anomaly, so that additional massless fields are required to cancel the anomaly. 

The most famous and well studied way to cancel the anomaly is to introduce $(26-D)$ additional massless bosons, which leads to the critical string theory.
Alternatively, at any $D$ a single additional massless scalar may also do the job, provided it has the appropriate coupling to the Euler density on the worldsheet. This path leads to the conventional non-critical strings \cite{Polyakov:1981rd} (a linear dilaton background). 

Motivated by  the worlsheet axion it was proposed \cite{Dubovsky:2015zey} that yet another possible way to cancel the worldsheet particle production is
to introduce $(D-2)(D-3)/2$  massless pseudoscalar fields,  transforming as an antisymmetric tensor under the $O(D-2)$ group of transverse rotations of a long string. At $D=4$ this is equivalent to adding a single axion, and integrability fixes the value of the coupling constant in (\ref{S1a}) to 
\be
\label{QI}
Q_a=\sqrt{7\over 16\pi}\approx 0.373176\dots\;.
\ee
The agreement with the value (\ref{Ql}), which was previously deduced from the lattice data, is quite intriguing. 

Three possible explanations were put forward to explain this coincidence. First, it can be purely numerological and does not teach us anything about the underlying physics. Second, it might suggest that the worldsheet axion becomes massless in the planar limit, so that the worldsheet theory is integrable in the pure gluodynamics in the planar limit. This possibility has been studied on a lattice and is ruled out  \cite{Mike}. The axion mass approaches a constant non-zero value at  $N\to \infty$.

Finally, the third option, which is the basis of the ASA, is that this coincidence indicates that the worldsheet theory becomes integrable in the high energy limit. This would not be completely surprising---we started with an asymptotically free theory, so the only reason we may get something non-trivial in the high energy limit is that we took  the infinite $N_c$ limit first. One may still expect  the end result  to be somewhat simple (recall also that all bulk amplitudes turn zero after taking the infinite $N_c$ limit even at finite energy). Note that if this interpretation were correct, 
the resulting integrable scattering amplitudes have a very transparent physical meaning.
The $S$-matrix is completely determined by the two-particle ``shock wave" phase shift (neither annihilations nor reflections are present), which takes the form \cite{Dubovsky:2012wk},
\be
\label{eis}
e^{2i\delta(s)}=e^{i\ell_s^2s/4}\;.
\ee
This phase shift results in a time delay proportional to the collision energy. This is the most basic geometric property of a relativistic string---its physical length is proportional to the energy, resulting in the time delay.

An antisymmetric tensor does not carry any local degrees of freedom at $D=3$, so in this case there is no need to add any massless states to restore integrability, and the phase shift (\ref{eis}) is compatible with a non-linearly realized bulk Poincar\'e symmetry for a theory of a translational Goldstone boson alone. However, as we already mentioned, lattice data exhibits sizeable corrections to the integrable phase shift both for $N_c=3$  \cite{Dubovsky:2014fma} and at large $N_c$ (the corresponding lattice data was recently published in \cite{Athenodorou:2016kpd} and its TBA analysis is currently underway and will be presented elsewhere). 

The ASA offers a unified description of the worldsheet dynamics for both $D=3$ and $D=4$ gluodynamics. Namely, the 
expectation is that in both cases the particle content of the worldsheet theory coincides with that for integrable axionic strings. At $D=4$ this implies that the worldsheet axion is the only additional massive degree of freedom, and at $D=3$ the particle content is just the same as in the classical Nambu--Goto theory.
A violation of integrability is a transient phenomenon present at intermediate energies. Integrability gets restored both in the IR (this is an automatic consequence of the non-linearly realized Poincar\'e symmetry) and in the UV (this is an assumption, motivated by the numerical coincidence between (\ref{Ql}) and (\ref{QI})).

As discussed extensively in \cite{Dubovsky:2012wk}, the most interesting property of the phase shift (\ref{eis})  is that its asymptotic UV behavior
(dubbed ``asymptotic fragility") is not described by a conformal fixed point and matches many properties of a gravitational theory rather than of a conventional field theory.
Even if the worldsheet theory turned out to be different from (\ref{eis}) in the UV, it is extremely plausible that its UV behavior still cannot be described by a conventional conformal fixed point. 
One reason to suspect this is because in the presence of a fixed point one may expect restoration of the bulk Lorentz symmetry in the UV, which is at odds with unitarity. 
Furthermore, the rotational symmetry of the Goldstone modes gives rise to the Noether currents which cannot be separated into a holomorphic and antiholomorphic parts, which is not expected to happen for a non-free CFT (and there does not seem to be a good candidate free CFT to describe such a putative UV fixed point).
Furthermore, it looks hard to construct examples of non-trivial conventional 2D RG flows involving shift invariant non-compact bosons.

Note that all existing examples of non-integrable asymptotically fragile theories \cite{Dubovsky:2013ira} exhibit the integrable UV behavior (\ref{eis}).
These can be understood as  solvable deformations of conventional quantum field theories by an irrelevant operator $\bar{T}T$ \cite{Smirnov:2016lqw,Cavaglia:2016oda} (see \cite{McGough:2016lol} for a 3D holographic description).
 Hence, it will be even more interesting if the expectation that the UV fixed point is absent does hold,
but UV integrability is not present, because this will provide a more general example of the asymptotic fragility. 

More generally,  this discussion gives an additional incentive to study the worldsheet dynamics of  confining large $N_c$ gauge theories.
This is likely to be a setting where lattice simulations combined with the large $N_c$ limit (which can also be taken numerically) provide an operational non-perturbative definition of (two-dimensional) gravitational theories.

Before moving forward, let us note the following. In the past a number of phenomenological flux tube models aimed to describe (parts of)
the glueball spectrum were proposed (such as \cite{Isgur:1984bm,Kaidalov:1999yd}).  Also a number of attempts have been made to calculate certain properties of the hadronic spectrum within the effective strings framework (e.g., \cite{Baker:2002km,Hellerman:2013kba,Sonnenschein:2015zaa}). The ASA is different from both. Unlike phenomenological flux tube models it has a very precise meaning---it represents (a guess for) the full set of quantum numbers and the structure of the spectrum of integrable axionic strings (in the short string sector; no guess is needed to describe the long string sector using the TBA). Within the ASA the integrable axionic strings are expected to provide a good zeroth order approximation for the planar Yang--Mills glueballs.  Unlike effective string models the ASA is not restricted to the long string regime. Effective strings provide a possible way to organize a perturbative expansion around the integrable approximation. Hopefully, after the integrable solution is understood, it may be possible to find other perturbative schemes as well, which will not be limited to long strings.

Finally, let us comment on a possible resolution of the apparent conflict between the simplicity of the ASA and the arguments put forward in \cite{Polchinski:1991tw}. As mentioned in the Introduction these arguments suggest that the worldsheet theory necessarily carries an infinite number of degrees of freedom. It looks plausible that the contradiction arises due to non-commutativity of the large $N_c$ and high energy limits.
Namely, the reasoning in \cite{Polchinski:1991tw} proceeds by compactifying  a gauge theory on a small thermal circle
of size $\beta$, such that the theory becomes perturbative, calculating the free energy (in the unstable phase with the unbroken center symmetry) and comparing the result with the analytic continuation of the calculation in the worldsheet theory. However, for the gauge theory to be in a perturbative regime the circle size should be smaller than $\beta<(\Lambda_{QCD}N_c)^{-1}$, which corresponds to the masses of the lightest off-diagonal Kaluza--Klein $W$-bosons arising as a result of compactification \cite{Unsal:2008ch}. At any finite value of $N_c$ this pushes the compactification scale above the UV cutoff (\ref{Luv}), where the worldsheet theory looses unitarity. Consequently,  the regimes of applicability of the two calculations never overlap, which may explain the discrepancy. In other words, this discrepancy may be taken as an indication that the perturbative asymptotically free regime is not accessible in the string description.

\section{Quantum Numbers of Closed Bosonic Strings and Glueballs}
\label{sec:spectrum}
In principle, the ASA hypothesis can be tested directly by further lattice simulations in the long string sector. In practice though it is rather 
challenging. Even with an improved convergence of the TBA technique the precision measurements of the worldsheet $S$-matrix require data for relatively large compactification radii. This makes it hard to measure many new states, because these become very heavy. 
On the other hand, a large number of high precision glueball mass determinations are available, so it would be very useful
to be able to confront the ASA against these data. Also, predicting the glueball spectra is what one normally means by solving the theory in the first place.

An important implication of the results reviewed in section~\ref{sec:long}, is that they set the scale of how ambitious one should be in ``solving" the theory. Indeed, a priori, one could aim at obtaining relatively simple exact analytic expressions for all glueball masses in the planar gluodynamics. However, the absence of integrability on the string worldsheet strongly suggests that this would be an overly ambitious goal. It appears more realistic to aim at understanding  the general structure of the glueball spectrum (in particular, at predicting glueball quantum numbers), and at developing a perturbative scheme(s) for calculating the masses.

In the present paper we restrict our ambitions even further, and aim only at understanding the quantum 
numbers. Even this limited approach will provide us with a powerful check of the ASA and of the stringy nature of the glueball states more generally. Namely, as we will argue now, the ASA allows to predict quantum numbers of almost all glueball states.
We will mostly restrict our discussion to the 3D case, both due to its relative simplicity and because the amount and quality of the 4D lattice data is much more limited. Nevertheless, a generalization to 4D appears straightforward and we will take a brief look at the 4D case as well.

\subsection{Tensor Square Structure}
\label{sec:tensor}
3D glueballs are characterized by their integer spins $J$, and $P$- and $C$-parities. The massive little group in three dimensions is $O(2)$, so that all states with non-zero spins come in degenerate pairs with opposite $P$-parities. So we will label $J>0$ states as $J^C$ with $C=\pm$.
(Pseudo)scalar states do not need to come in pairs, so we will label them as $0^{PC}$ with $P,C=\pm$.

A first implication of the ASA, and even more generally of the closed string nature of glueball states, is the following general structure of the  Hilbert space. Let us start with the simplest case, when no massive states are present on the worldsheet, which is the situation in the 3D ASA. Then, a (non-chiral) closed string is an object with two identical sets of left- and right-moving excitations. So one may expect the  Hilbert space of closed string states to have a natural tensor square structure. This is indeed the case in the long (winding) string sector,
\be
\label{longtensor}
 {\cal H}_{long}={\cal H}_L\otimes {\cal H}_R\;,
 \ee
 where ${\cal H}_{long}$ is the Hilbert space of all possible excitations on the worldsheet of a long string, and ${\cal H}_{L(R)}$ are (identical) Hilbert spaces of left(right) moving excitations.

However, as well known from the critical string theory,  allowed states for short strings have to satisfy an additional condition called level matching. This condition has a very general physical origin and is not limited to critical strings. It is a manifestation of the worldsheet reparametrization invariance, which implies that the total momentum along a string has to vanish. This does not hold in the long (winding) string sector, where translations along the string are generated by a combination of the worldsheet translation (which is a gauge symmetry transformation) and of the target space translation (which is a global symmetry transformation). As a result, for long strings states with a non-zero spatial 
momentum along the winding  direction the level matching condition does not hold.
Instead, for short strings one finds the following structure of the Hilbert space,
\be
\label{shorttensor}
{\cal H}_{closed\; short}=\sum_{N} {\cal H}_L(N)\otimes {\cal H}_R(N).
\ee
Here $N$ is a ``level" of a state, so that
 ${\cal H}_{L(R)}(N)$ is a subspace of left(right) moving states with a total momentum $N$ along the string. We choose the spatial worldsheet coordinate to run in the range $\sigma\in [0,2\pi]$, so that $N$ takes natural values. 

It is straightforward to generalize the structures (\ref{longtensor}), (\ref{shorttensor}) to the situation when massive fields are present on the worldsheet, as relevant for the 4D ASA. In this case additional zero momentum states are present, corresponding to adding $n$ massive 
particles at rest. For instance, for short strings one gets then 
\be
\label{shortmassive}
{\cal H}_{massive}=\sum_n {\cal H}_n\;,
\ee
where each of ${\cal H}_n$ has the structure of (\ref{shorttensor}).

Clearly, to test the structure (\ref{shorttensor}) against the actual glueball data measured on a lattice, one needs some assumption about the distribution of glueball masses. For critical (integrable) strings all states at the same level are degenerate in mass. An approximate integrability of the ASA makes it then natural to expect that glueballs should come in approximately degenerate groups ordered by their level.
%In fact, as we will see soon, even a somewhat weaker assumption will be enough to turn the structure (\ref{shorttensor}) into a useful test of a stringy nature of glueballs.
 
 Then the immediate implication of the structure  (\ref{shorttensor}) is that glueball states come in clusters with multiplicities given by squares of integers.  Let us see now what are the possible  quantum numbers inside these clusters.
 First, note that the spin $J$ and the spatial parity $P$ can be defined at the level of left- and right-moving states separately. On the other hand, the charge conjugation $C$ is directly linked to the tensor product structure of the closed string Hilbert space. Indeed, in a gauge theory the
 charge conjugation reverses the direction of path ordered exponents, used to construct glueball states. 
 In the string language this translates into a spatial parity transformation on a worldsheet,  {\it i.e.}, $C$ acts by exchanging quantum numbers of left- and right-moving components of a closed string state. 
 
 Then in the decomposition of (\ref{shorttensor}) we encounter the following terms,
 \begin{gather}
 \label{same0}
 0^{P}\otimes 0^{P}=0^{++}\\
 0^{P_1}\otimes 0^{P_2}+0^{P_2}\otimes 0^{P_1}=0^{(P_1P_2)+}+0^{(P_1P_2)-}
 \label{different0}\\
 0^{P}\otimes J+J\otimes 0^{P}=J^++J^-
 \label{0J}\\
 \label{sameJ}
 J\otimes J = (2J)^{+}+0^{++}+0^{--}\\
 \label{differentJ}
 J_1\otimes J_2+ J_2\otimes J_1=(J_1+J_2)^++(J_1+J_2)^-+|J_1-J_2|^++|J_1-J_2|^-\;.
 \end{gather}
 One consequence of these relations is that the leading Regge trajectory---{\it i.e.}, the set of  maximum spin states at each level---consists of states of even spin. Furthermore, odd spin states necessarily come in pairs of opposite $C$-parity, because they may only appear in a decomposition   of  products of different spins, eqs.~(\ref{0J}) and (\ref{differentJ}). 
 %Finally, as we find in section \ref{sec:spins}, 
 %we will encounter only products of the type $0^{+}\otimes 0^{+}$ in the scalar sector. Hence, yet another general prediction of the ASA is that all spin 0 glueballs are either $0^{++}$ or $0^{--}$.
 
\subsection{Spin Content}
\label{sec:spins}
The very existence of the tensor structure (\ref{shorttensor}) already implies  a quite restrictive set of constraints on the glueball spectrum. However, as we will argue now, 
with a couple of additional assumptions,
the ASA implies a much stronger set of predictions and allows to almost completely fix the set of glueball quantum numbers at each level. The first major additional assumption will be that at each level the Hilbert spaces describing left- and right-moving excitations ${\cal H}_{L(R)}(N)$ have the same structure as the level $N$ Hilbert space of open strings,
\be
\label{Hlro}
{\cal H}_L(N)= {\cal H}_R(N)= {\cal H}_{open}(N)\;.
\ee
This assumption holds for critical strings, and given that the worldsheet $S$-matrix is the same for critical strings and for  integrable  axionic strings, it appears as a natural one to make. The main a posteriori justification for this assumption, as discussed in section~\ref{sec:3Data}, is coming from the lattice data.  

To describe the full open string Hilbert space, let us present it as a sum of subspaces corresponding to different spin eigenvalues,
\[
{\cal H}_{open}\equiv \sum_N{\cal H}_{open}(N)=\sum_{J\in {\mathbf Z}}{\cal H}_J\;.
\]
At large $|J|$ the minimum energy configuration in each  ${\cal H}_J$ subsector may be described as a classical rotating rod solution of the Nambu--Goto theory,
\begin{gather}
\label{X0}
X^0={\ell^2_sE\over \pi}\tau\\
\label{Xrod}
X\equiv X^1+iX^2={\ell_s\over 2i}\sqrt{2 J\over \pi}\l \e^{i\sigma^+}-\e^{i\sigma^-}\r\;,
\end{gather}
where we picked $J>0$, which is related to the energy $E$ by the classical Regge formula
\[
E^2={2\pi J\over \ell_s^2}\;.
\]
In (\ref{Xrod}) we introduced the worldsheet light cone coordinates
\[
\sigma^\pm\equiv \tau\pm \sigma
\]
 with $\sigma\in [-\pi/2,\pi/2]$. Then the structure of  ${\cal H}_J$ at large $J$ can be described 
perturbatively by quantizing small excitations around the rotating rod solution.
 Moreover, at large $J$ one may ignore the conformal anomaly and use the classical Polyakov formalism. 
 We work in the conformal gauge and fix the residual gauge freedom by imposing that $X^0$ stays in the form (\ref{X0}) (where $E$ becomes a total perturbed energy now). Then, restricting to perturbations in the rest frame of the rotating rod, we can write
 \be
 \label{modes}
 \delta X=i{\ell_s\over\sqrt{\pi}}\sum_{n\neq 0}{\alpha_n \over 2n}\l \e^{-in\sigma^-}\e^{in\pi/2}+\e^{-in\sigma^+}\e^{-in\pi/2}\r\;,
 \ee
where the commutation relations are
\[
[\alpha_m,\alpha^\dagger_n]=2m\delta_{m-n}\;,
\]
so that the conventionally normalized creation operators are
\[
a_n^\dagger={\alpha_{-n}\over \sqrt{2n}}\;,\;\; b_n^\dagger={\alpha^\dagger_{n}\over \sqrt{2n}}\;,\;\;n>0\;.
\]
In addition to the usual Virasoro constraints
\be
\label{Vir}
-(\d_+X^0)^2+\d_+X\d_+\bar{X}=-(\d_-X^0)^2+\d_-X\d_-\bar{X}=0
\ee
 the states belonging to ${\cal H}_J$ have to satisfy the condition that their angular momentum is equal to $J$, {\it i.e.} 
 \be
\label{J}
{1\over 2 i\ell_s^2}\int_{-{\pi\over 2}}^{\pi\over 2}d\sigma\l\bar{X}\d_\tau X-X\d_\tau\bar X\r=J
\ee
By plugging in the mode decomposition (\ref{modes}) into the angular momentum constraint at linear order in the perturbation $\delta X$ one obtains,
\be
\delta J=-i\sqrt{\pi J\over 2}(\alpha_{-1}^\dagger-\alpha_{-1})=0\;.
\ee
On the other hand, at linear order in $\delta X$, $\delta E$ the Virasoro constraints imply the following relations
\begin{gather}
\delta E=-{i\pi\over 2 \ell_s}(\alpha_{-1}^\dagger-\alpha_{-1})\\
\alpha_{-2}=0\\
\label{Ln}
\alpha_{-n}=\alpha^\dagger_{n-2}\;,\;\; n\geq 3
\end{gather}
We see that at the linear order 
\[
\delta E=0
\] 
at constant $J$, as it should be because the rotating rod solution minimizes the energy at  fixed angular momentum. We also see that the Virasoro constraints completely remove the $(\alpha_{-2},\alpha_{-2}^\dagger)$ sector (which is related to the fact that we consider dynamics in the rest frame of the rotating rod). Furthermore, after imposing the $\delta J=0$ constraint, the only remaining solution in the   $(\alpha_{-1},\alpha_{-1}^\dagger)$ 
sector corresponds to a non-consequential constant time shift. 

From the worldsheet viewpoint each of the remaining Virasoro constraints (\ref{Ln}) restricts the motion  in the space of two harmonic oscillators with frequencies equal to $n$ and $n+2$ by imposing
\be
\label{Lna}
\sqrt{n+2}a^\dagger_{n+2}=\sqrt{n}b^\dagger_n\;\;\;n>0\;.
\ee
This constraint appears somewhat unconventional. Naively, one would like to say that it projects out one of the 
oscillators. However, given that before the projection the oscillators had different frequency it is somewhat confusing what should be the frequency of the remaining one.
The physical meaning of this condition is much cleaner from the target space-time viewpoint.
Indeed, the physical time translations are generated by a simultaneous shift of the worldsheet time $\tau\to\tau+\tau_0$ and a rotation in the
$(X_1,X_2)$ plane, $X\to\e^{-\tau_0} X$.
Operators $a^\dagger_{n+2}$ and $b^\dagger_n$ carry $\pm 1$ unit of charge under the rotation, so from the space-time point of view the constraint is imposed in the phase space of two oscillators of equal  frequency
\be
\label{frequencies}
\omega_n={\pi (n+1)\over \ell_s^2 E}\;,
\ee
leaving us with a single oscillator of the frequency  (\ref{frequencies}) for each $n>0$.

To confirm this interpretation by a direct calculation let us inspect the leading (second order) expression for the perturbed energy. Expanding the corresponding Virasoro constraint to the second order we obtain
\be
\label{dE2nd}
\delta E={i\pi\over 2 \ell_s}(\alpha_{-1}^\dagger-\alpha_{-1})+{\pi\over \ell_s^2 E}\sum_{n>0}\l(n+2)a_{n+2}^\dagger a_n+n b_n^\dagger b_n\r\;,
\ee
where we kept the $(\alpha_{-1}^\dagger-\alpha_{-1})$-term because it does not vanish at the second order. Indeed, the angular momentum constraint expanded to the second order becomes
\[
\delta J=i\sqrt{\pi J\over 2}(\alpha_{-1}^\dagger-\alpha_{-1})+\sum_{n>1}\l a_{n+2}^\dagger a_{n+2}- b_n^\dagger b_n\r=0\;.
\]
By plugging this back into the expression for the energy (\ref{dE2nd}) we obtain
\be
\delta E={\pi\over \ell_s^2 E}\sum_{n>0}(n+1)\l a_{n+2}^\dagger a_{n+2}+ b_n^\dagger b_n\r
\ee
in agreement with the previous discussion.
Now it is straightforward to impose the constraint (\ref{Lna}).
%
%
%
%
%The leading (quadratic in perturbations) expression for the energy can be determined by keeping ${\cal O}(\delta X^2)$ terms in the Virasoro constraint and reads as
%\[
%\delta E={1\over \ell_s^2}\int_{-{\pi\over 2}}^{\pi\over 2}d\sigma\d_+\delta X\d_+\delta\bar X=
%{1\over \ell_s^2}\int_{-{\pi\over 2}}^{\pi\over 2}d\sigma\d_-\delta X\d_-\delta\bar X\;.
%\]
%
Namely, after defining a rotated set of oscillators 
\begin{gather}
A_n={\sqrt{n}a^\dagger_{n+2}+\sqrt{n+2}b_n^\dagger\over \sqrt{2(n+1)}}
\\
B_n={\sqrt{n+2}a^\dagger_{n+2}-\sqrt{n}b_n^\dagger\over \sqrt{2(n+1)}}
\end{gather}
the constraint (\ref{Lna}) turns into 
\[
B_n=0\;,
\]
and the energy reduces to
\be
\delta E={\pi\over \ell_s^2 E}\sum_{n>0}(n+1)A_n^\dagger A_n
\ee

\begin{figure}[t!!]
	\centering
	\includegraphics[width=0.75\textwidth]{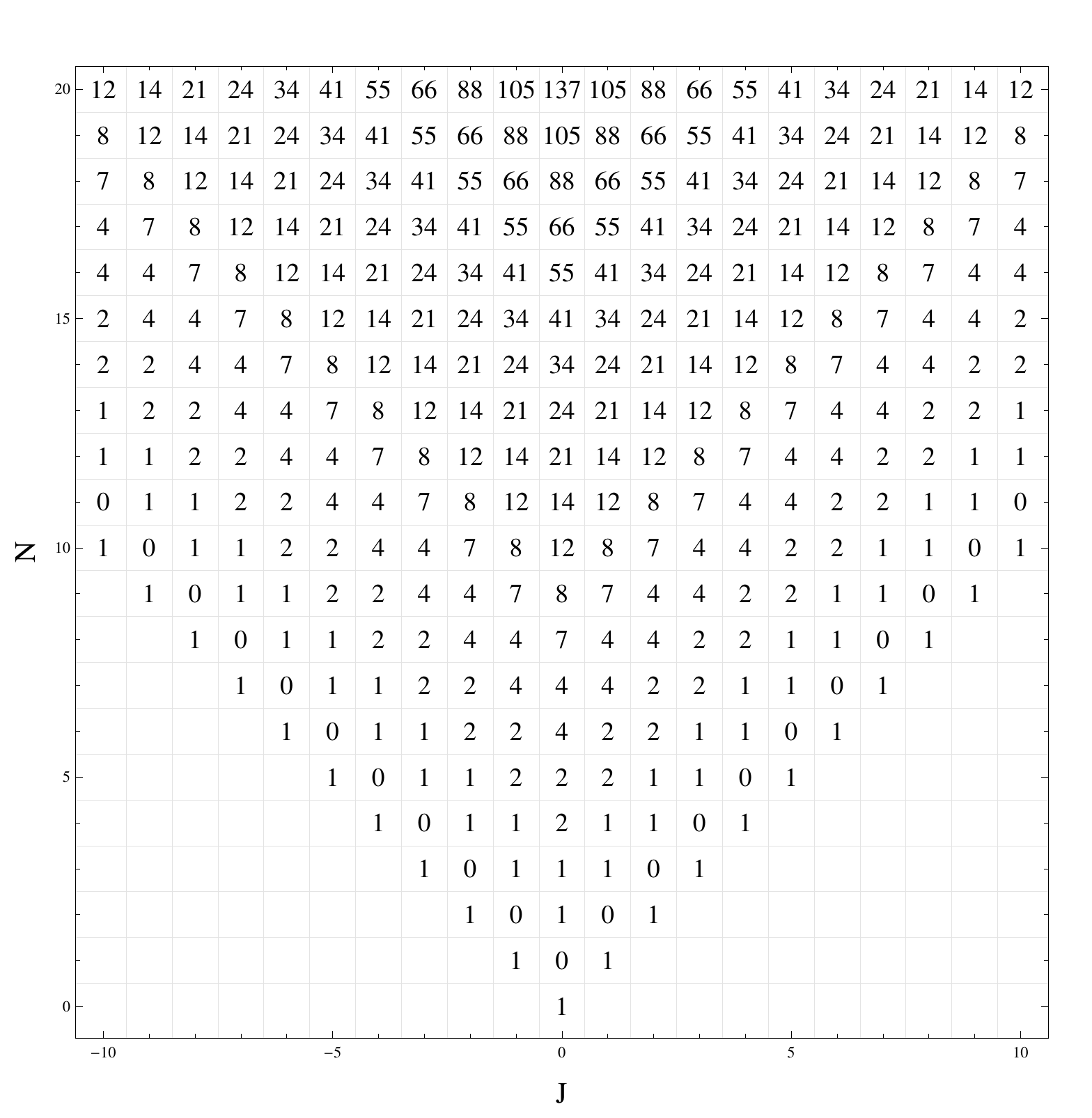}
	\caption{Helicity content of the ASA open string spectrum.}
		\label{fig:PASA}
\end{figure}

To proceed, let us keep following the assumption that the structure of level assignments in ${\cal H}_J$ is the same as for critical strings. 
Then the ground state in ${\cal H}_J$ corresponds to the level $N=J$, and each oscillator $A_n^\dagger$ increases the level by the amount determined by its frequency, {\it i.e.}  by
\[
\Delta N_n=n+1\;.
\]
Hence, the multiplicity of states at a level $J+N$ in ${\cal H}_J$ is equal to the number of partitions of the number $N$, which do not include 1 (because the set of frequencies in (\ref{frequencies}) does not include 
$\omega_0$). This can be summarized by the following generating function
\be
\label{PJ}
P_J(x)=x^J(1-x){\cal P}(x)\;,
\ee
where
\be
{\cal P}(x)=\prod_{n=1}^\infty {1\over 1-x^n}
\ee
is the Euler generating function for integer partitions. Then the multiplicities of states at different levels in $\cal{ H}_J$ can be read off from the Taylor expansion of (\ref{PJ}),
\be
\label{TaylorPJ}
P_J(x)=x^J+x^{J+2}+x^{J+3}+2x^{J+4}+2x^{J+5}+4x^{J+6}+4x^{J+7}+7x^{J+8}+\dots\;.
\ee
In section~\ref{sec:Higgs} we will confirm this result by comparing it to the standard bosonic string 
character \cite{Curtright:1986di} (see Fig.~\ref{fig:Plc}).

It is straightforward to generalize this analysis to $D=4$, where additional set of oscillators, describing perturbations in the transverse direction, are present. Also it is straightforward to include additional massive states on the worldsheet.

The perturbative analysis above is only guaranteed to work in the semiclassical limit $J\to \infty$. However, it is natural to try to ``analytically continue" these results all the way down to $J=0$. Namely, we include as a part of the ASA the assumption that the generating function (\ref{TaylorPJ}) (or its appropriate generalization at $D=4$) describes the structure of ${\cal H}_J$ for any $J$.  This is the second major assumption. It is probably more questionable than the first one (given by (\ref{Hlro})). As before, its main justification is that it seems to be supported by data.

Then the structure of the full open string Hilbert space ${\cal H}_{open}$ can be characterized by the following generating function
\be
\label{ourP}
P_{ASA}(x,\theta)\equiv\sum_{J\in {\mathbf Z}}\e^{iJ\theta}P_J(x)={(1-x)(1-x^2){\cal P}(x)\over 1+x^2-2x\cos\theta}
\ee
The coefficient in front of $x^N\e^{ih\theta}$ in the Taylor expansion (in $x$) of $P_{ASA}(x,\theta)$ is equal to the multiplicity of the helicity $h$ open string states at the level $N$. In Fig.~\ref{fig:PASA} we present the resulting multiplicities for a number of low lying levels.
\begin{table}[t!]
\begin{center}
 \begin{tabular}{ | c  | c | c | }
   \hline			
 $N$   & Glueball States &  $\#$ of states\\
    \hline
    0 & $0\otimes 0=0^{++}$ & 1\\
    \hline  
    1 & $1\otimes 1=0^{++}+0^{--}+2^+$ & 4\\
    \hline
    2 & $(0+2)\otimes(0+2)
    =2\cdot 0^{++}+0^{--}+2^++2^-+4^+$ & 9\\
    \hline
     3 & $(0+1+3)
     \otimes(0+1+3)
     =$ & 25\\
     &  $3\cdot 0^{++}+2\cdot 0^{--}
     +1^++1^-+2\cdot 2^++2^-+3^++3^-+4^++4^-+6^+$&\\
    \hline
   \end{tabular}
  \caption{Quantum numbers of 3D glueballs at first four levels, as predicted by the ASA.}
      \label{tab:3Dspectrum}
  \end{center}
\end{table}

The last remaining step to obtain the 3D glueball spectrum, as predicted by the ASA, is to calculate tensor squares at each level, as explained in section~\ref{sec:tensor}. 
Note that the analytic continuation above apparently does not
 allow to determine the parity assignment for $J=0$ states in ${\cal H}_{open}$. However, as follows from (\ref{same0})-(\ref{differentJ}), this ambiguity may only affect the glueball spectrum at $J=0$ and in a situation when more than one state is present in ${\cal H}_{open}$ at $J=0$. This does not happen until $N=4$, which is beyond what is currently accessible on a lattice.
 The resulting predictions for glueball quantum numbers up to $N=3$ are presented in Table~\ref{tab:3Dspectrum}.
\section{Comparison to Lattice Data}
\label{sec:lattice}
We are now in a good position  to confront the predictions of the ASA with the glueball spectrum measured on a lattice. We will mainly consider the 3D case, which is simpler both on the theory side and as far as lattice simulations go.
In particular,  the quality  of lattice data is much better at $D=3$ than at $D=4$. However, we will take a brief look at $D=4$ glueball spectra as well.
\subsection{3D glueballs versus ASA}
\label{sec:3Data}
\begin{figure}[t!!]
	\centering
	\includegraphics[width=\textwidth]{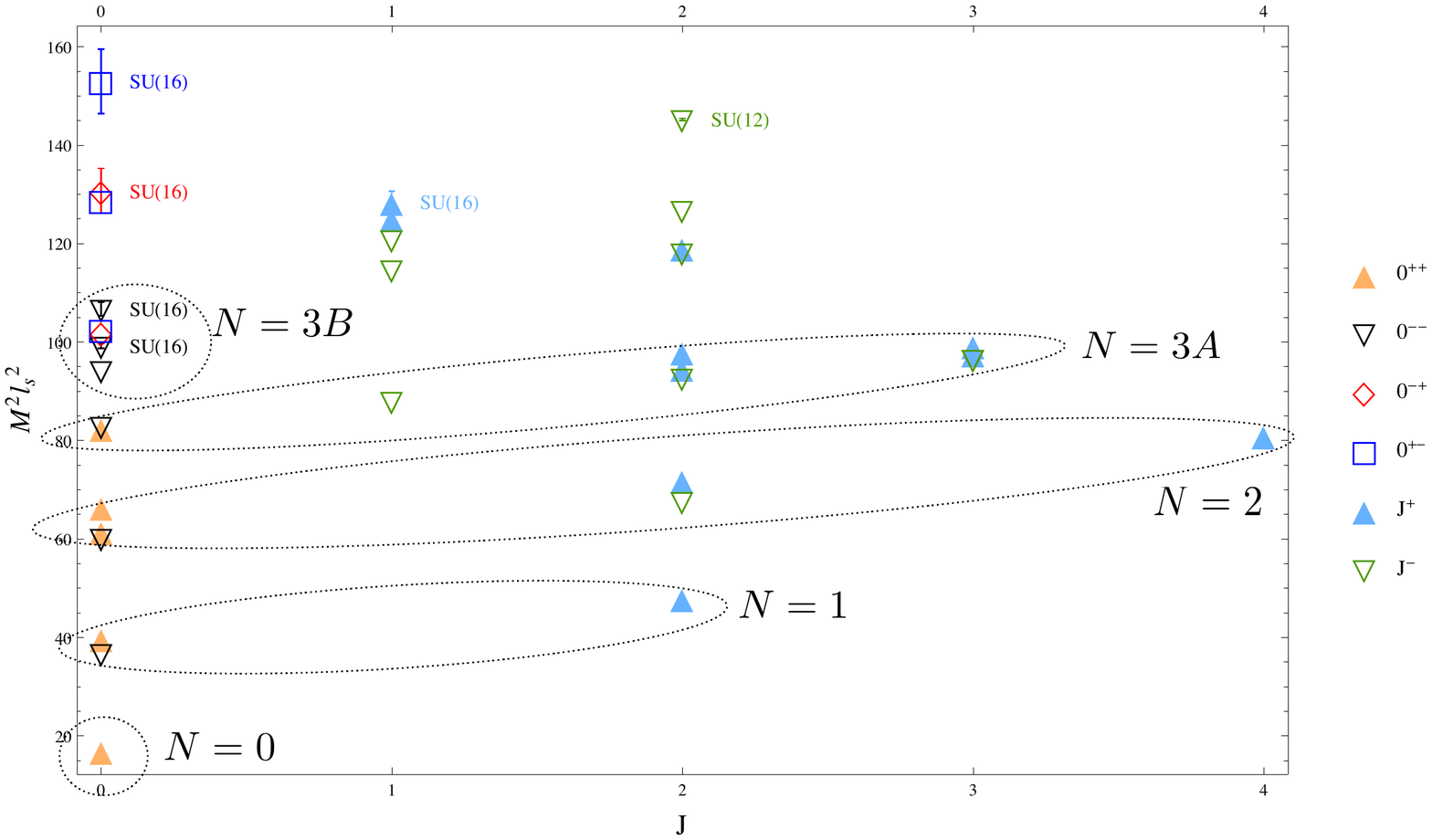}
	%{glueball_spectrum_3d_original.pdf}
	\caption{3D glueball spectrum and quantum numbers as measured on a lattice.}
		\label{fig:raw3D}
\end{figure}

An  extensive high precision determination of glueball spectra in $D=3$  gluodynamics at large $N_c$ has been reported recently in  \cite{Athenodorou:2016ebg}. In Fig.~\ref{fig:raw3D} we present the resulting glueball spectrum. Several comments are in order to properly interpret this data.

Most importantly, one needs to keep in mind that a square lattice breaks the $O(2)$ rotation symmetry down to its ${\mathbf Z}_4$ subgroup.
This makes it rather challenging to determine the actual spins of glueball states. Without additional efforts one only determines a ${\mathbf Z}_4$ representation corresponding to a given spin. This implies that all spins divisible by 4, $J=0\,\mbox{\rm mod}\,4$,  are coming out indistinguishable, as well as all odd spins, $J=1\,\mbox{\rm mod}\,2$, and all
even spins non-divisible by 4, $J=2\,\mbox{\rm mod}\,4$.
These degeneracies may be resolved with special dedicated techniques \cite{Meyer:2002mk,Meyer:2004gx}. However, currently these have been implemented only for a handful of states, and the results should be considered as preliminary. Namely, spin determinations were performed for spin 3 and spin 4 states shown in Fig.~\ref{fig:raw3D}.
Also low lying $J=0$ states do not have a candidate partner of opposite parity and have to be scalars under continuous rotations.
However,  one should keep in mind that some of the states labeled as spin 0, 1 and 2 may actually have a higher spin.

Another point to make is that Ref.~\cite{Athenodorou:2016ebg} presented results  for $SU(N_c)$ gauge groups with $N_c=2,3,4,6,8,12,16$ as well as extrapolations to $N_c=\infty$. 
Most of the states shown in Fig.~\ref{fig:raw3D} are taken from the $N_c=\infty$ Table of  Ref.~\cite{Athenodorou:2016ebg}. However, 
for a few states the mass determination at large $N_c$ was not accurate enough to allow for a high quality $N_c=\infty$ extrapolation. These states were not included in $N_c=\infty$ results in \cite{Athenodorou:2016ebg}, even though they are present either in $N_c=12$ or $N_c=16$ data.
Given that our current analysis   mostly depends
on having as complete set of states as possible, and does not rely much on the precise mass determination, we included such states in Fig.~\ref{fig:raw3D}, with a subscript indicating what is the largest number of colors ($N_c=12$, or $N_c=16$), where a state is present in \cite{Athenodorou:2016ebg}.

Finally, for the same reason and to avoid an unnecessary cluttering of the plot, we did not include mass error bars in Fig.~\ref{fig:raw3D} for the $N_c=\infty $ states from  \cite{Athenodorou:2016ebg}. All these mass determinations are at a few percent level precision, which is more than enough for our purposes. For states which lack $N_c=\infty$ extrapolation, we took the average mass between the two highest values of $N_c$ where these are  measured as a central value, and the corresponding mass difference as an error bar.

To compare the spectrum in Fig.~\ref{fig:raw3D}  with the ASA expectations, let us first ignore the results of section~\ref{sec:spins} and see how much can be deduced just from imposing the tensor square structure alone. This may be considered as the most basic test of the stringy nature of the glueball states. 

A separation between different levels is quite manifest in Fig.~\ref{fig:raw3D}, at least for low lying glueball states.
% In Fig.~\ref{fig:raw3Dl}, the same 
%data is presented together with how we combine glueballs into different levels.
 It is immediately clear that the level degeneracy is only approximate. In fact, the level structure is better visible at fixed value of $J$ (with caveats concerning  spin determinations, as just discussed). If we were to completely suppress the spin information it would be much harder to sort the states between different levels. Also,
 it definitely appears that  higher spin states tend to be heavier at any fixed level number.

Nevertheless, there is a clearly separated lightest $0^{++}$ state, which is natural to identify with a single glueball expected at $N=0$ level. Next we find a well separated group of four states, $0^{++}$, $0^{--}$, and
$2^{+}$ (recall that the $2^{+}$ glueball, as well as all other $J\neq 0$ glueballs, is actually a doublet of two opposite parity states).
These four states can be uniquely written as the tensor square,
\be
0^{++}+0^{--}+2^+=1\otimes 1\;.
\ee
 It is natural to identify these with the $N=1$ level, which matches nicely the ASA prediction at the first level as presented in Table~\ref{tab:3Dspectrum}.

Moving higher in mass we find a group of nine states corresponding to $2\cdot 0^{++}$, $0^{--}$, $2^+$, $2^-$, and $4^+$ glueballs. Again these can be uniquely written as a tensor square,
\be
2\cdot 0^{++}+0^{--}+2^++2^-+4^+=(0+2)\otimes(0+2)
\ee
in a perfect agreement with the ASA decomposition of the $N=2$ level in Table~\ref{tab:3Dspectrum}. 

At even higher masses we encounter the first and the only major contradiction between the observed glueball spectrum and the tensor square structure. Namely, as we discussed in section~\ref{sec:tensor}, the tensor square structure requires that odd spin states come in pairs with opposite $C$-parities. This is violated by the spectrum presented in Fig.~\ref{fig:raw3D}. One does find there a group of four approximately degenerate odd spin states, of which two have $C=+$ and the other two have $C=-$. However, there is a single $J=1$ (and $C=-$) state in this group, which is incompatible with (\ref{same0})-(\ref{differentJ}). So, to be able to proceed, we assume that the spin assignment for one of the odd spin states in this group is misplaced. Verifying this assignment becomes crucial not only for falsifying the ASA, but as a test of the tensor square structure, which is a much more basic consequence of the stringy nature of glueballs.

Even with this assumption, identifying the $N=3$ states is not immediate. Indeed, there is a well pronounced group of sixteen states, labeled as 3A in Fig.~\ref{fig:raw3D}. However, among these states one finds seven $C=-$ states and nine $C=+$ states. This indicates the 3A states on their own cannot be presented as a tensor square (otherwise the split would be six $C=-$ states and ten $C=+$ states).

At this point one might entertain a possibility that sixteen 3A states should be combined with nine $N=2$ states; this would imply that we actually misidentified what represents the $N=2$ level. A priori, this were not completely unreasonable.  For instance, the $J=0$ 3A states are quite close in mass to the $J=4$, $N=2$ state. However, even though the counting of $C$-parity would work in this case (we would get ten $C=-$ states and fifteen $C=+$ states, as needed), it is immediate to see that this option does not work. Indeed, the best shot at presenting the resulting  twenty five states as a tensor square would be to consider the product $(0+1+2)\otimes(0+1+2)$, which  would predict six odd spin glueballs instead of four, which are present among 3A states.

Hence, the only remaining option to reconcile the spectrum in Fig.~\ref{fig:raw3D} with the tensor square structure is to combine the 3A states with a somewhat heavier group of $J=0$ states, labeled as 3B.  This association looks more convincing after consulting with the  ASA predictions
in Table~\ref{tab:3Dspectrum}. These imply that many of the 3B states are actually $J=4$ states, and we already saw that higher spin states at the same level tend to be heavier.  Namely,  both $0^{-+}$ and $0^{+-}$ 3B states are predicted to be members of $4^+$ and $4^-$ doublets. Then it is natural to expect  that the $0^{--}$ state, which is nearly degenerate with the $0^{+-}$, is actually the second member of $4^{-}$.
The only remaining rearrangement needed to become compatible with the ASA at $N=3$ is to identify the heaviest $2^+$ state as  $6^+$.

\begin{figure}[t!!]
	\centering
	\includegraphics[width=\textwidth]{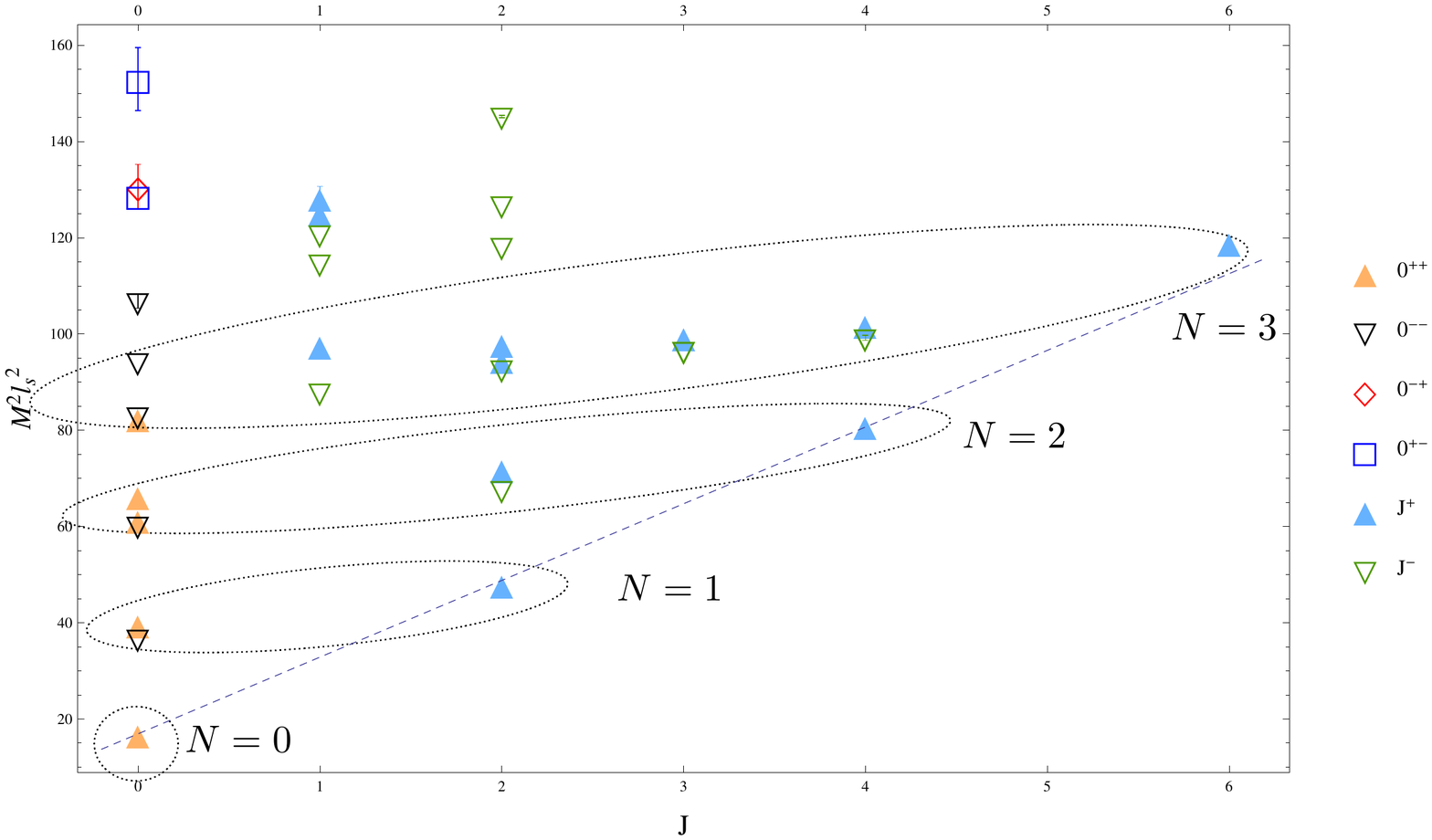}
	\caption{Measured 3D glueball spectrum with the ASA quantum numbers assignments. The dashed line shows the approximately linear leading Regge trajectory.}
		\label{fig:3Dour}
\end{figure}

The resulting glueball spectrum is shown in Fig.~\ref{fig:3Dour}. Comparing it to Table~\ref{tab:3Dspectrum}, we observe that to make them compatible three additional $0^{++}$ states need to be present (one of which is actually the second component of $4^+$). 
Most likely all these states are heavier than the  $0^{++}$ states found on the lattice so far, so their existence is not in conflict with the results of \cite{Athenodorou:2016ebg}.

To summarize, we believe that the agreement between the ASA predictions for the first four levels and the lattice results is quite encouraging. The only contradiction between the two is that the ASA requires one of the states identified as $3^+$ in \cite{Athenodorou:2016ebg} to be actually $1^+$. Such a misidentification is not completely implausible\footnote{We thank Mike Teper for encouraging us to proceed with the ASA in spite of this contradiction.}, and testing the spin determination for this pair of states becomes very important. In addition, the ASA makes a number of predictions for the currently undetermined spins and predicts the existence of three additional $0^{++}$ states  in the mass range corresponding to $N=3$ level in Fig.~\ref{fig:3Dour}.
 
Clearly, there is not enough lattice data at higher masses to make a meaningful comparison with the ASA predictions
at $N=4$. Let us nevertheless take a brief look at what is expected to happen there. The ASA predicts the following sixty four states at $N=4$ level,
\begin{gather}
\label{N4}
(0^{P_1}+0^{P_2}+1+2+4)\otimes(0^{P_1}+0^{P_2}+1+2+4)=5\cdot 0^{++}+3\cdot 0^{--}+
\\
\nonumber
+0^{P_1P_2+}+0^{P_1P_2-}+3\cdot (1^++1^-)+4\cdot 2^++3\cdot 2^-+2\cdot(3^++3^-)+3\cdot4^++2\cdot 4^-+\\
\nonumber
+5^++5^-+6^++6^-+8^+
\end{gather}
At this level for the first time we encounter a situation where quantum numbers of some glueballs remain undetermined by the ASA. However,  we see that this affects only two states out of sixty four, and there are only two physically distinct options,
$P_1P_2=\pm 1$.  
It will be interesting to test these predictions with the future lattice data.
\subsection{A Brief Look at 4D Glueballs: Worldsheet Axion}
\label{sec:4D}
Comparing the ASA predictions to lattice simulations in four dimensions is more challenging both due to larger uncertainties in lattice glueball spectra, and because the ASA structure is more involved now, with an additional massive excitation present on the worldsheet. For this reason we defer a detailed 4D analysis  for a separate publication. Here we will take only a brief look at the $D=4$ spectra. By comparing them with the tensor square structure we will see that the data indeed  suggests the presence of a massive  worldsheet axion, in agreement with the earlier long string analysis and with the ASA.

Note first that the massive little group in four dimensions is $O(3)$, so that the spatial parity $P$ commutes with rotations. As a result, unlike in 3D, all glueball multiplets are labeled by $J^{PC}$.
Then in the decomposition of the tensor square expression (\ref{shorttensor}) we will encounter the following terms,
\begin{gather}
\label{4DdifferentJ}
J_1^{P_1}\otimes J_2^{P_2}+J_2^{P_2}\otimes J_1^{P_1}=\sum_{J=|J_1-J_2|}^{J=J_1+J_2}\l J^{(P_1P_2)+}+J^{(P_1P_2)-}\r\\
\label{4Dsame}
J^P\otimes J^P=\sum_{0\leq k\leq 2 J}(2J-k)^{+(-1)^k}\;.
\end{gather}
As we saw in the 3D case the very existence of the tensor square structure imposes a quite restrictive set of requirements on the glueball spectrum. Let us see whether these can be met at $D=4$.

\begin{figure}[t!!]
	\centering
	\includegraphics[width=\textwidth]{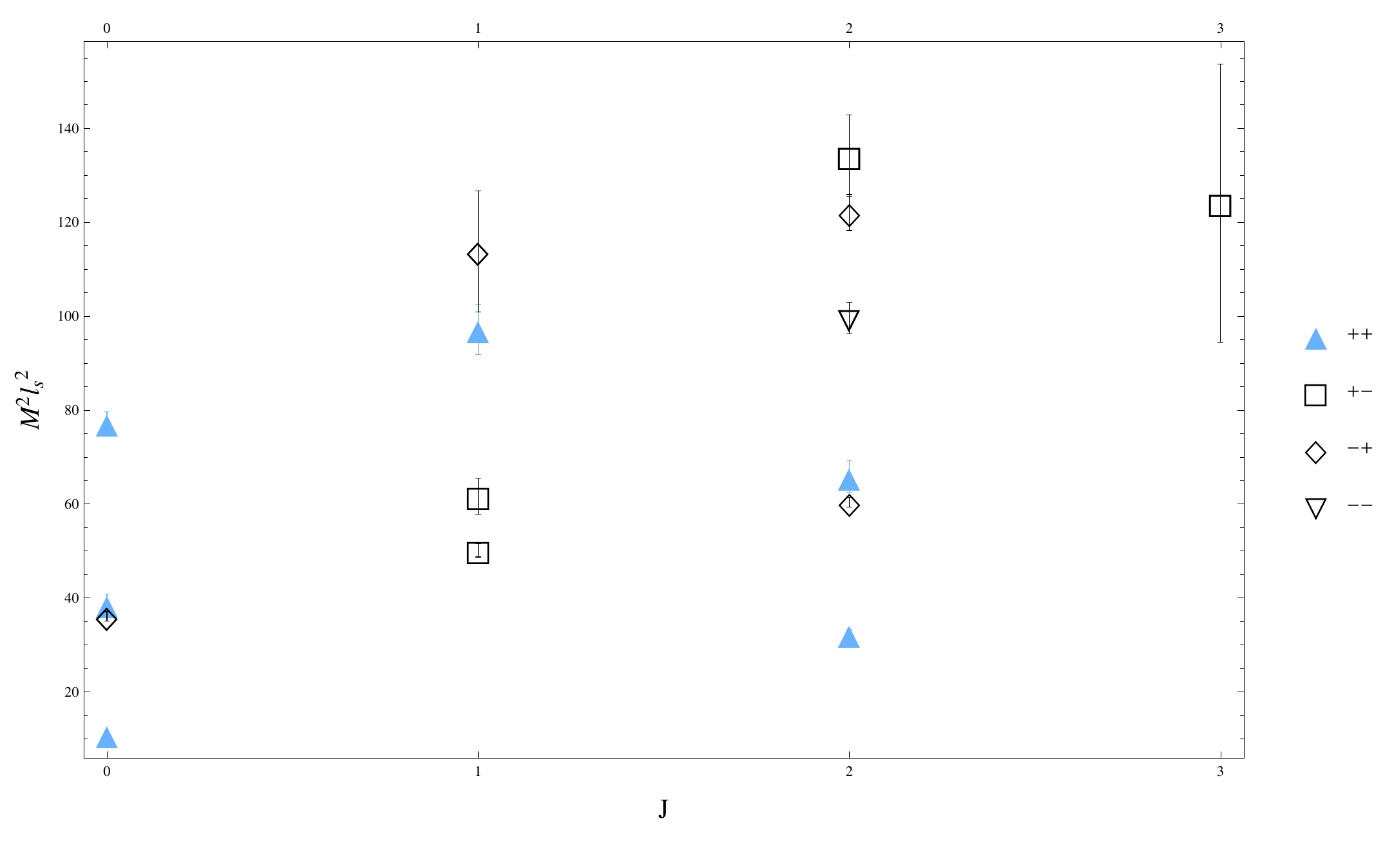}
	\caption{4D glueball spectrum and quantum numbers.}
		\label{fig:4dglueballs}
\end{figure}
 In Fig.~\ref{fig:4dglueballs} we presented the large $N_c$ spectrum of low lying glueballs as measured in \cite{Lucini:2010nv}. 
Just as in 3D one immediately identifies the lightest $0^{++}$ glueball as the only $N=0$ state. However, already at $N=1$ level one encounters a problem. Namely, the natural $N=1$ candidates  are the first excited $0^{++}$ glueball and the lightest $0^{-+}$, $1^{+-}$ and $2^{++}$ glueballs. In total this gives $2+3+5=10$ states, which is incompatible with the tensor square structure. Nevertheless, according to (\ref{4Dsame}), nine of these states do provide a tensor square
\be
1^{P}\otimes 1^P=0^{++}+1^{+-}+2^{++}\;,
\ee
independently of $P$.
Also, as follows from (\ref{4DdifferentJ}),  (\ref{4Dsame}), to obtain a $0^{-+}$ state within a tensor square structure would require to include quite a large number of additional states at $N=1$ level, and in particular $0^{--}$, in a clear disagreement with the measured spectrum.
This leaves us with two options. First, one might conclude that the lightest $0^{-+}$ glueball does not fit into a string interpretation at all. 

Alternatively, one may argue that this state should be thought of as the $N=0$ state with an additional zero momentum massive $P=-$, $C=+$ excitation added on the worldsheet. The latter interpretation is exactly what one expects from the ASA. Indeed, the worldsheet axion identified in \cite{Dubovsky:2013gi} is a pseudoscalar both under reflections in a transverse plane to the long string and with respect to the spatial parity on its worldsheet. The former is the $P$-parity of the gauge theory, and the latter is the combination of $P$ and $C$, so adding such an excitation to the $N=0$ $0^{++}$ state may indeed transform it into $0^{-+}$. Note, however, that this interpretation leaves unexplained a very near degeneracy between $0^{-+}$ and the first excited $0^{++}$.

\section{Noncritical Strings and the Higgs Mechanism}
\label{sec:Higgs}
We feel that the results of the previous section provide a considerable support for the ASA, especially in 3D. Thus it appears quite plausible that 3D glueballs may be described by a closed bosonic string theory without any extra
local degrees of freedom on the worldsheet. This conclusion definitely needs to be supported by a further quantitative analysis of the glueball mass spectrum. However, one may wonder whether the qualitative analysis presented  here sheds any light on the decades old puzzles, which had been  preventing  a construction of bosonic string theory of the QCD string. The goal of the present and the next section is to initiate such a discussion. In the current section we focus on the problems present at any number of dimensions different from $D=26$, and in the next section we discuss specific $D=3$ puzzles.

To start with, let us stress that at any $D$ it is straightforward to develop an effective low energy theory describing perturbations around  long strings \cite{Luscher:1980ac,Luscher:2004ib,Aharony:2010db,Aharony:2011ga,Dubovsky:2012sh,Aharony:2013ipa}, or around rotating strings with a high angular momentum, similar to how we proceeded in section~\ref{sec:spins} (see, e.g., \cite{Baker:2002km,Hellerman:2013kba,Sonnenschein:2015zaa}). This is not what we are after here;  our goal is to build a theory describing all string states, including the shortest ones. Then a possible starting point may be a light-cone quantization  \cite{Goddard:1973qh} of the Nambu--Goto theory.
At least it produces a consistent unitary quantum theory on the worldsheet, which is not so bad as a start. However, one immediately encounters the following three major problems.
\begin{itemize}
\item The lightest short string state has a negative mass squared---{\it the tachyon problem}.
\item{\it Breaking of the target space Poincare symmetry} at a non-critical number of dimensions, $D\neq 26$.
\item{\it Massless photon (graviton)} is present in the spectrum of open (closed) strings.
\end{itemize}
Note that the second and the third problems are closely related. To see this, let us focus on the open string case; 
as before, closed strings are best understood as a tensor square of the open ones. 

The light cone quantization is not manifestly covariant, so the resulting glueball states come out in multiplets of the group of transverse rotations $O(D-2)$, rather than of the full massive little group $O(D-1)$. However, somewhat surprisingly, at all $N\neq 1$ levels  the resulting $O(D-2)$ multiplets can be combined into complete representations of $O(D-1)$ \cite{Curtright:1986di}. Instead, at $N=1$  one finds a single $O(D-2)$ vector.
Hence, the simplest way to arrive at a fully covariant theory is to require that the $N=1$ state is a massless gauge boson. This fixes the value of the normal ordering constant $a=1$ and $D=26$ and leads to the three problems above. 

This discussion makes it clear that resolving the three problems is akin to finding an analogue of the Higgs mechanism. At $N=1$ level additional states are required, 
which would allow to make the $N=1$ vector massive, in turn opening up the possibility for the tachyonic $N=0$ state to be lifted into the $m^2>0$ region as well.
However, in general, the string dynamics does not allow to introduce states at the $N=1$ level only.

To make this general discussion more concrete let us illustrate how it all works  using an example of the simplest non-critical string theory---the linear dilaton CFT.
In the language used in section~\ref{sec:long} the linear dilaton comes from an integrable worldsheet theory of a long string with an additional {\it scalar} degree of freedom \cite{Dubovsky:2015zey}. Integrability is restored thanks to the interaction of the form
\be
Q\int d^2\sigma\sqrt{-g}R\phi\;,
\ee
where 
\[
Q=\sqrt{25-D\over 48 \pi}\;.
\]
To describe the corresponding short string spectrum it is convenient to make use of the observation that at the quantum level this setup is equivalent to starting with a classical Nambu--Goto theory in $D+1$ dimensions, and employing the light-cone quantization \cite{Daszkiewicz:1997ax}. Of course, the linear dilaton spectrum can also be studied using the standard worldsheet CFT approach.

Then, in the sector with zero momentum along the dilaton ``direction", the resulting short string spectrum can be described in the following way. At $N=0$ one finds a scalar tachyon, as before. All higher level states can be constructed by acting on the tachyon state with a set of creation operators $A^{I\dagger}_n$ transforming as a vector under the massive little 
group $O(D-1)$ for any $n$, so that $I=1,\dots D-1$. Here $n=1,2,\dots$ and the level of a state is equal to
\[
N=\sum n_i\;,
\]
where $n_i$ are the oscillators corresponding to the state.
Restricting to $D=3$, we can then summarize the spin content of  linear dilaton strings by the generating function similar to (\ref{ourP}),
\be
\label{Pld}
P_{ld}={\cal   P}(x\e^{i\theta}){\cal   P}(x\e^{-i\theta})=\prod_{n=1}^\infty{1\over 1+x^{2n}-2 x^n \cos\theta}\;.
\ee
This spectrum is manifestly covariant and there appears to be no obstruction to deform the theory away from integrability in such a way that the tachyon mass gets lifted into the $m^2>0$ region. Introducing the dilaton mass (potential) looks like a natural step in this direction, which would also justify setting to zero the momentum in the dilaton direction. The resulting deformed spectrum will exhibit then also additional states corresponding to adding a massive dilaton at rest, as in (\ref{shortmassive}).
 
 The fastest way to see a direct parallel between the linear dilaton setup and the Higgs mechanism is to recall that at any $D$ the spin content of the minimal light-cone quantized bosonic string $P_{lc}$ is related to the linear dilaton generating function as \cite{Curtright:1986di}
 \be
 \label{Plc}
 P_{ld}={P_{lc} {\cal P}(x)}\;.
 \ee
 So linear dilaton adds to the minimal light cone spectrum a missing $O(D-2)$ singlet state at $N=1$ to form an $O(D-1)$ vector,
 \[
 \yng(1)_{D-1}
 =\yng(1)_{D-2}+singlet_{D-2}\;,
 \]
and, as a consequence of string dynamics, the $N=1$ singlet brings in a tower of singlet states at higher levels.
This path appears very similar to the conventional field theoretic Higgs mechanism. However, as we saw, this is not what happens for confining strings. There is no sign of a scalar massive dilaton state neither in the $D=3,4$ spectra of long string, nor in the spectra of glueballs. Also for a massless linear dilaton theories in $D=3$ level multiplicities following from $P_{ld}$ (1, 4, 25, 100, ... for closed strings) are vastly larger than the observed ones, which agree well with the ASA (1, 4,  9, 25,...).

If anything, the matter content of the ASA is suggestive of another field theoretic way to make a vector massive. Instead of ``eating" a scalar, an $O(D-2)$ vector can get ``eaten" itself by 
an $O(D-2)$ antisymmetric tensor to form an $O(D-1)$ antisymmetric tensor,
\[
\yng(1,1)_{D-1}=\yng(1)_{D-2}+\yng(1,1)_{D-2}\;.
\]
It remains to be seen whether this is the right way to think about ASA.
Definitely, it gets somewhat confusing at $D=3$, where an antisymmetric tensor of the massive little group $O(2)$ is equivalent to a scalar, while the massless little group $ O(1)$ reduces to 
$P$-parity only. 

In fact, even coming back to the linear dilaton, the meaning of (\ref{Plc}) is somewhat ambiguous  at $D=3$. Indeed, at $D>3$ the helicity content of a light-cone quantized string can be deduced by analyzing the set of $O(D-2)$ multiplets at each level. For $D=3$, this does not work due to triviality of $O(1)$. Instead, at $D=3$ one may look at (\ref{Plc}) as a definition of $P_{lc}$ by analytic continuation in $D$ from $D>3$. Recalling the expression (\ref{Pld}) for $P_{ld}$ we obtain at $D=3$,
\[
P_{lc}={{\cal   P}(x\e^{i\theta}){\cal   P}(x\e^{-i\theta})\over {\cal P}(x)}=\prod_{n=1}^\infty{1-x^n\over 1+x^{2n}-2 x^n \cos\theta}\;.
\]
It is natural to try to come up with a similar interpretation for $P_{ASA}$, as given by (\ref{ourP}). However,
dividing  $P_{ASA}$ by $P_{lc}$ one observes that the simplest generalization of (\ref{Plc}) fails---the $P_{ASA}/P_{lc}$ ratio does not result in a generating function with positive Taylor coefficients.

\begin{figure}[t!!]
	\centering
	\includegraphics[width=0.75\textwidth]{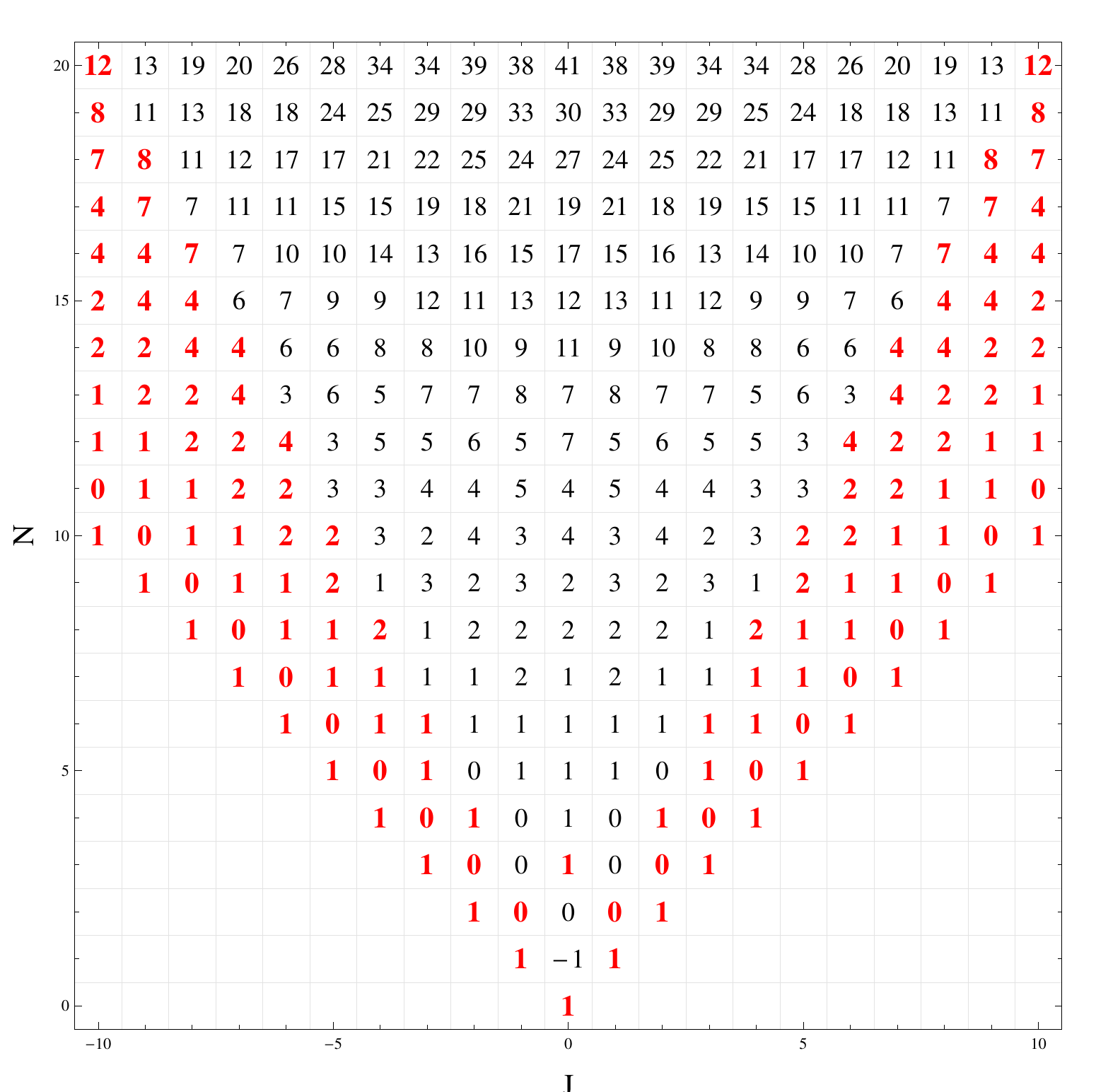}
	\caption{Helicity content of the $D=3$ light cone open string spectrum. State multiplicities which agree with the corresponding ASA values are shown in bold red.}
		\label{fig:Plc}
\end{figure}
To get a further insight into how $P_{lc}$ and $P_{ASA}$ are related, in Fig.~\ref{fig:Plc} we presented the helicity content for a number of low lying levels for $P_{lc}$ similarly to how we previously did for $P_{ASA}$ in  Fig.~\ref{fig:PASA}. We see that at $N=1$ level a single negative coefficient appears as a result  of taking the ratio $P_{ld}/P_{lc}$. This is an indication that the $N=1$ states do not form complete $O(2)$ multiplets, as discussed earlier.

As Fig.~\ref{fig:Plc} demonstrates, ASA and light cone multiplicities
agree for sufficiently low lying states at each value of $J$ up to some maximum level $N_{max}(J)$, which becomes larger and larger as $J$ grows. This is expected, because this region corresponds to a semiclassical regime where the light cone quantization is trustworthy and should agree with the semiclassical ASA spectrum. At $N>N_{max}(J)$ we observe that there are always more states in the ASA spectrum than in the light cone one, suggesting that some version of the Higgs mechanism is still at play here. Furthermore, if one suppresses the helicity content and inspects the sheer multiplicities of states at different levels one finds the relation
\be
\label{ASAlc}
P_{ASA}(x,\theta=0)=(1+x)P_{lc}(x,\theta=0)\;,
\ee
which is suggestive of the presence of a non-local  fermionic worldsheet degree of freedom in the 3D ASA.
\section{Puzzles of 3D Strings}
\label{sec:anyons}
According to the ASA the worldsheet theory of confining strings in 3D  does not carry any local degrees of freedom in addition to the minimal bosonic strings. This provides a good motivation to discuss several puzzling features of bosonic strings at $D=3$\footnote{In fact, most of the present discussion extends to superstrings as well.}. 

There are several approaches to string dynamics, which apparently lead to conflicting conclusions at $D=3$.
First, there is the conventional Polyakov treatment. At first sight the $D=3$ case does not appear special at all in this language. There is a non-vanishing conformal anomaly in the path integral over the worldsheet metric, which makes the Liouville mode dynamical. 

The situation is very different when one applies the light cone quantization to $D=3$ strings \cite{Mezincescu:2010yp}. At general $D$ the Poincar\'e anomaly in the light cone quantization manifests itself in the following non-vanishing commutator,
\[
[J^{-i},J^{-j}]\neq 0\;.
\] 
This commutator is proportional to $(D-26)$ and vanishes for critical strings.
However, as was emphasized in  \cite{Mezincescu:2010yp},  this commutator also vanishes  trivially at $D=3$, because  $i=j$ in this case.
This led to the proposal that the light cone quantization might be consistent at $D=3$. The resulting spectrum depends on a single continuous parameter---the normal ordering constant $a$, which determines the string intercept. For critical strings $a=1$ from requiring the closure of the Poincar\'e algebra, but it remains undetermined at $D=3$. Independently of the value of $a$ the resulting spectrum is anyonic---{\it i.e.}, contains states of irrational helicity. This spectrum was obtained by a direct calculation at $D=3$.
It is different from what we called the light cone spectrum in section~\ref{sec:Higgs}, which was obtained by the analytic continuation of the light cone character from higher dimensions. The total multiplicities at each level are the same for both spectra and 
 given by the Euler partitions ${\cal P}(x)$.

Finally, $D=3$ strings also appear special from the viewpoint of long strings and integrability. Indeed, as we already discussed in section~\ref{sec:long},
the classical Nambu--Goto string is integrable at any $D$. At one loop the integrability is broken and the resulting non-elastic amplitudes are proportional to $(D-26)$. However, they also vanish at $D=3$ for trivial kinematical reasons, similarly to the anomalous commutator in the light cone quantization. Moreover, at any loop order the integrable $S$-matrix (\ref{eis}) is compatible with the non-linearly realized Poincar\'e symmetry without need for any extra local degrees of freedom on the worldsheet. In fact, this $S$-matrix can be derived as a unique integrable solution of the Poincar\'e Ward identities \cite{Dubovsky:2015zey}.

All this sounds similar to what one finds in the light cone quantization of short strings. However, there are also important differences.
First, in the long string sector the normal ordering constant $a$ is uniquely fixed by the worldsheet Poincar\'e symmetry (which is a consequence of the target space Poincar\'e symmetry). Given that $a$ is a UV dominated quantity it is natural to expect that it should be fixed in the short string sector as well, but the light cone quantization is somehow missing this.

Furthermore, 
one of the motivations for the ASA is that the $D=3$ integrable worldsheet theory is a member of a continuous family of axionic theories which exist for any value of $D$\footnote{Note, however, that the arguments presented in \cite{Dubovsky:2015zey} establish integrability of these theories at $D=3,4$ only. It remains to be seen whether integrability can be preserved at general $D$.}. This observation appears to be at odds with the presence of anyons---one does not expect to find irrational helicities at $D=3$ if analytic continuation in $D$ were possible.

Our expectation for the resolution of these apparent contradictions is that the standard light cone quantization performed in \cite{Mezincescu:2010yp} is to be considered anomalous at $D=3$ as well. The situation appears somewhat analogous to what happens in $SU(N_c)$ gauge theories with a single chiral quark generation in the fundamental representation. At $N_c>2$ such a theory is anomalous, and the anomaly shows up in local Ward identities. For $N_c=2$ the anomaly  is invisible at the level of local current algebra, but the theory is still anomalous at the non-perturbative level \cite{Witten:1982fp}. 

In the present context, the presence of anyons in the spectrum by itself is a strong indication of anomaly. The classical Nambu--Goto action is invariant under the Poincar\'e group $ISO(1,2)$ rather than under its universal cover, so one does not expect to find anyons in the spectrum if a quantization were to preserve the classical symmetry.
Note that in this sense the Nambu--Goto theory is different from the point particle case discussed in \cite{Mezincescu:2011nh}, where the Wess--Zumino term leads to irrational helicity already at the classical level.
 This does not necessarily imply that it is impossible to build a consistent interacting theory of anyons along the lines of  \cite{Mezincescu:2010yp}, although it would be natural to expect that the Wess--Zumino term allowed for 3D bosonic strings should play a role in such a theory \cite{Curtright:2010zz}. As a basic consistency test of the anyonic quantization it would be interesting to see whether it agrees with the ASA spectrum in the semiclassical regime, similarly to what we see in Fig.~\ref{fig:Plc} for $P_{lc}$.
  At any rate, even if such a consistent anyonic quantization were possible, this is not the path we are taking here. 

Let us see now that there is actually no conflict between the Polyakov formalism and the long string/integrability considerations.
Given the manifest covariance of the Polyakov formalism, this supports an idea that short 3D strings can be quantized in a way compatible with integrability and preserving the invariance under the Poincar\'e group rather than under its cover.

We will be using the version of the Polyakov formalism due to Polchinski and Strominger (PS) \cite{Polchinski:1991ax} (or rather its recent ``simplified" version presented in \cite{Hellerman:2014cba}).
In this formalism one writes the path intergal for a long bosonic string as
\be
\label{Polyakov}
Z=\int {\cal D}X{\cal D}h\; e^{iS_P+iS_{PS}+\dots}\;.
\ee
Here $S_{P}$ is the standard Polyakov action
\[
S_{P}=-{1\over 2\ell_s^2}\int d^2\sigma\sqrt{-h}h^{\alpha\beta}\d_\alpha X^\mu\d_\beta X_\mu
\]
and $S_{PS}$ is the PS action,
\[
S_{PS}=\beta\int d^2\sigma\sqrt{-h}\l h^{\alpha\beta}\d_\alpha \phi\d_\beta \phi-\phi R(h)\r
\]
with $\phi$ being any composite scalar operator transforming as the Liouville field under the Weyl transformations of the worldsheet metric.
The simplest choice is
\[
\phi=-{1\over 2}\log  h^{\alpha\beta}\d_\alpha X^\mu\d_\beta X_\mu\;.
\]
The coefficient $\beta$ is fixed by requiring that the Weyl anomaly of the integration measure in (\ref{Polyakov}) is cancelled by the classical Weyl variation of $S_{PS}$,
\[
\beta={D-26\over 24\pi}\;.
\]
Finally, dots in (\ref{Polyakov}) stand for non-universal terms, which are suppressed in the derivative expansion around the long string background. As we said,  the value $D=3$ does not appear special in this formalism at this point.

Let us see now how to derive the worldsheet $S$-matrix in this language. We start with the critical case $D=26$\footnote{We thank Juan Maldacena for suggesting the idea of this derivation.}. Alternatively, one may think of this step as calculating the leading order worldsheet amplitudes in a formal  ``$(D-26)$-expansion" around $\beta=0$.

We are interested in the scattering around the long string vacuum, so we write
\begin{gather}
X^0=\tau+ Y^0\\
X^{1}=\sigma+ Y^{1}
\label{Xalpha}
\\
X^i=Y^i\;,
\end{gather}
where $i=2,\dots,D-1$. In addition, let us fix the conformal gauge $h_{\alpha\beta}=\eta_{\alpha\beta}$.
In this case $Y^\mu$ solve a free flat space wave equation.
In particular, the solution for the transverse fields can be written as
\be
\label{free}
Y^i=\int_{-\infty}^\infty {dp \over\sqrt{ 2\pi}}{1\over\sqrt{ 2E}}\l \alpha^{i\dagger}_{p}e^{i (E\tau -p\sigma)}+h.c. \r\;,
\ee
where $E=\pm p$. 
 $Y^\alpha$ ($\alpha=1,2$) are determined by the Virasoro constraints.  
 
 Naively, the solution (\ref{free}) corresponds to a free $S$-matrix.
However, the $S$-matrix (\ref{eis}) describes the experience of observers which use physical target space coordinates  $X^0$, $X^1$ rather than the worldsheet ones. This is similar to what we saw in section~\ref{sec:spins}, where the physical meaning of the constraints (\ref{Lna}) becomes transparent only after switching to frequencies w.r.t. the physical time $X^0$. 
Hence to calculate the worldsheet $S$-matrix we need to rewrite (\ref{free})
in the form
\be
\label{Xfree}
Y^i=\int_{-\infty}^\infty {dp \over\sqrt{ 2\pi}}{1\over\sqrt{ 2E}}\l a_{out(in)}^{i\dagger}(p)e^{i (EX^0 -pX^1)}+h.c. \r\;,
\ee
in the two asymptotic regions $X^0\to \pm\infty$, corresponding to $a_{out(in)}^{i\dagger}$.

Let us switch to the light cone coordinates both on the worldsheet and in the target space. Then the Virasoro constraints can be written as
\begin{gather}
Y^-(\sigma^+,\sigma^-)={}\int_{-\infty}^{\sigma^+}d\tilde{\sigma}^{+}
\l \d_+Y^i\d_+ Y^i- \d_+ Y^+\d_+ Y^-\r+y^-(\sigma^-)\;,
\label{Ym}\\
\label{Yp}
Y^+(\sigma^+,\sigma^-)={}\int_{-\infty}^{\sigma^-}d\tilde{\sigma}^{-}
\l \d_- Y^i\d_- Y^i-\d_- Y^+\d_- Y^-\r+y^+(\sigma^+)\;.
\end{gather}
We may choose
\be
\label{y0}
y^+(\sigma^+)=y^-(\sigma^-)=0\;.
\ee
This is equivalent to imposing a version of the light cone gauge condition, which is well suited for the long string sector, 
\be
\label{LC}
\d_- Y^-=\d_+Y^+=0\;,
\ee
Then the second term under the integrals in (\ref{Ym}), (\ref{Yp}) vanishes as well. As a result we obtain in the two asymptotic regions,
\begin{gather}
Y^\pm(\tau\to -\infty)=0\\
Y^\pm(\tau\to\infty)={\ell^2_s\over 2} {\cal P}^\pm\;,
\end{gather}
where ${\cal P}^\pm$ are the worldsheet momenta operators of transverse string modes. 
Using this in (\ref{Xfree}) to express $\tau,\;\sigma$ through $X^0,\;X^1$ we obtain that 
\begin{gather}
a^{i\dagger}_{in}(p_\pm)=\alpha^{i\dagger}(p_\pm)\\
a^{i\dagger}_{out}(p_\pm)=e^{-i\ell_s^2 p^\pm {\cal P}^\mp/4}\alpha^{i\dagger}(p_\pm)\;
\end{gather}
Then for a two particle state we get
\[
a^{i\dagger}_{in}(p^+)a^{j\dagger}_{in}(p^-)|0\rangle=e^{i\ell_s^2 p^+ {\cal P}^-/4}a^{i\dagger}_{out}(p^+)e^{i\ell_s^2 p^- {\cal P}^+/4}a^{j\dagger}_{out}(p^-)|0\rangle=e^{i\ell_s^2s/4}a^{i\dagger}_{out}(p^+)a^{j\dagger}_{out}(p^-)|0\rangle\;,
\]
reproducing (\ref{eis}). In the same way we can directly calculate the scattering phase shift for an arbitrary multiparticle process.

Let us turn now to the non-critical case and calculate the leading in $\beta$ corrections to the above $S$-matrix.
At leading order in $\beta$ these can be read off from the PS interaction obtained by setting
$Y^\pm$ to their values determined from the leading order Virasoro constraint, so that one obtains the following vertex,
\be
\label{directPS}
S_{PS}=-{2\beta}\int d\sigma^+d\sigma^-\d_+\phi\d_-\phi\;,
\ee
where
\[
\phi=-{1\over 2}\log\l 2-\d_+Y^i\d_-Y^i+{1\over 8}(\d_+Y^i)^2(\d_-Y^j)^2\r\;.
\]
To obtain multiparticle amplitudes at leading order in $\beta$ one just needs to Taylor expand (\ref{directPS}) and read off $n$-particle amplitude from ${\cal O}(Y^n)$ term of the resulting series. It is straightforward to check that at ${\cal O}(Y^4)$ one reproduces the one-loop annihilation amplitude calculated in \cite{Dubovsky:2012sh}, and at ${\cal O}(Y^6)$ one obtains inelastic one-loop six particle amplitudes calculated in 
\cite{Cooper:2014noa}. It is a rather remarkable property of the Nambu--Goto action that one can write down a 
simple generating functional for all one-loop amplitudes, which are nevertheless non-trivial.
 
Finally, we see that $D=3$ is special in the Polyakov formalism as well. All these universal amplitudes, which imply the breakdown of integrability at $D\neq 26$, vanish on-shell at $D=3$, because for a single flavor $Y$ all interaction vertices of the form $\d_+(\d_+Y)^n\d_-
(\d_-Y)^n$ are trivial on-shell. So as expected, the Polyakov formalism agrees with the static gauge analysis of the worldsheet scattering.

Of course, this is only a first perturbative step towards deriving the integrable $D=3$ worldsheet theory 
in the Polyakov formalism. To complete the derivation one needs to show, similarly to the static gauge analysis of \cite{Dubovsky:2015zey},
that higher order terms in (\ref{Polyakov}) can be chosen such that integrability persists at all orders in the derivative expansion.

It will be very interesting to see whether the Polyakov formalism allows to build a tractable path integral description of integrable $D=3$ strings. Indeed, the analysis based on the $S$-matrix (\ref{eis}) allows to calculate the finite volume spectrum in the long string sector through TBA. However, at least in its present form, it does not allow to calculate the short string spectrum. The Polyakov language is manifestly covariant and equally well suited for both sectors. Note, that normally one interprets the path integral (\ref{Polyakov}) as an effective description valid only for  soft excitations around semiclassical string configurations. A plausible expectation is that in the microscopic theory the Polchinski--Strominger interaction is coming from integrating out a massive Liouville mode. We see that at $D=3$ this is not the only possibility and the theory can be defined as a microscopic one (at least in the long string sector) without introducing any new local degrees of freedom and without ``resolving" the PS interaction.

To avoid a confusion let us stress that the above analysis does {\it not} imply that the PS interaction can be dropped or removed by a field redefinition in the path integral (\ref{Polyakov}). 
In particular, it contributes to the central charge of the worldsheet conformal theory.
Related to this, in agreement with   \cite{Polchinski:1991ax}, we do not expect that the PS  theory can be promoted to a well-defined conformal field theory at $D=3$. Rather, the PS interactions vanish on-shell when restricted to physical string states (cohomologies of the corresponding BRST operator in the conformal gauge\footnote{In our streamlined derivation of the worldsheet scattering we avoided introducing the FP ghosts and the use of the BRST formalism for the sake of simplicity. However, the BRST approach appears the only responsible way to fully describe the integrable $D=3$ string in the conformal gauge.}), and
the $D=3$ theory may be made well-defined only when restricted to physical observables (at least in the long string sector).
The PS interaction 
 is crucial for the consistency of the path integral  (\ref{Polyakov}) (in particular to ensure the nilpotence of the BRST operator in the conformal gauge). It is natural to expect it will play a role in calculating the anyon-free short string spectrum, if such a calculation is at all possible within this approach.
 
In fact, we expect that  to define $D=3$ strings in the short string sector a {\it non-local} worldsheet degree of freedom is required, which is invisible in the long string sector. We already saw an indication for this at the end of the previous section, when comparing $P_{ASA}$ and $P_{lc}$, (\ref{ASAlc}). To expand on that argument, note that naively one might expect left-moving level $N$ states ${\cal H}_L(N)$ appearing in the tensor square formula (\ref{shorttensor}) to be isomorphic to momentum $N$ left-movers in the winding sector, so that
\[
{\cal H}_L=\sum_N{\cal H}_L(N)\;,
\]
where ${\cal H}_L$ appears in (\ref{longtensor}). We see from (\ref{ASAlc}) that this is not the case for the ASA string. Level multiplicities
in the winding string sector are given by integer partitions ${\cal P}(x)$, so that (\ref{ASAlc}) signals that there are ``more" left-movers in the short string sector than for the long string. Something analogous happens for NSR strings where for short strings there is an extra freedom in choosing boundary conditions for fermions, which is invisible for an infinitely long string. In the present context one may expect that the extra non-local fermionic degree of freedom suggested by  (\ref{ASAlc}) is related to ``framing" of knots in the Chern--Simons theory \cite{Witten:1988hf}.

\section{Conclusions}
\label{sec:last}
We think the most solid conclusion from the presented analysis is that glueballs in the large $N_c$ gluodynamics walk, talk and quack as excitations of a bosonic string (at least at $D=3$). We presented a dynamical Ansatz---the ASA---for the structure of the corresponding bosonic string theory. Already at the present qualitative level this leads to a number of predictions for glueball quantum numbers, which can be tested on a lattice. Especially important at this stage will be to have spin determinations for a larger number of glueball states.

Of course, if all qualitative ASA predictions are confirmed, one would like to move further and to calculate quantitative properties 
of the ASA spectrum. Ideally, this involves first solving for the spectrum of short strings in the integrable approximation, and then developing a perturbation theory around the integrable spectrum. In fact, at large spin this program can be pushed forward even without knowing the exact  solution within the effective string theory approach along the lines of \cite{Baker:2002km,Hellerman:2013kba,Sonnenschein:2015zaa}. 
Clearly there are a number of challenges which need to be solved to implement this program. For instance,  the existing perturbative calculations of the leading closed string Regge trajectory  do not apply directly at physical dimensions $D=3,4$, due to endpoint singularities of the corresponding semiclassical string solution (a folded rotating rod).
Let us report here a  little numerical  puzzle, which might be related to this problem. In Fig.~\ref{fig:3Dour} one observes what looks as a  nice approximately  linear leading Regge trajectory passing through the lowest energy $J=0,2,4,6$ glueball states.
It can be approximated by a straight line (the dashed line in Fig.~\ref{fig:3Dour}),
\be
\label{1Regge}
\ell_s^2 M^2\approx 1.27\cdot  4\pi( J+1.06)\;.
\ee
We see that the slope of this line is significantly (by a factor of 1.27) steeper than what is expected from the classical rotating folded rod solution. It well may be that this discrepancy disappears at higher $J$, although with the current data we see no tendency for the slope to get flatter as $J$ grows. One should keep in mind though that the $J=6^+$ state on this Regge trajectory is the heaviest measured glueball state in its sector (shown as $J=2^+$ in Fig.~\ref{fig:raw3D}, because the spin determination has not yet been performed for this state). So it might be that a somewhat lighter state is missing in the lattice data in this sector.
Of course, any of these explanations  would imply that the observed approximate alignment of the four points is to some extent coincidental. On the other hand, this discrepancy seems to go in line with the tendency for the mass to grow with $J$ at each level. 

Another intriguing numerological property of the Regge trajectory (\ref{1Regge}) is that after factoring out the overall $1.27\cdot 4\pi$ coefficient,
its intercept is quite close to unity.
Resolving these puzzles may well be the first necessary step towards quantitative stringy description of the glueball spectrum.

Coming back to general questions about the ASA, note that the proposal consists of two parts. First there is an assumption about the matter content on the worldsheet. In addition, 
 there is a guess, inspired by the long string data, that the UV worldsheet asymptotics is integrable. It is fair to say that the analysis presented here mostly tests the first ASA assumption. It is an interesting open question what is the best set of physical observables, which would allow to probe the UV asymptotics of the worldsheet scattering. The spectrum of low lying glueballs is mostly sensitive to the momenta of order $\ell_s^{-1}$. However, heavy glueballs are also not immediately useful for this purpose due to the UV/IR feature of string excitations---most of the heavy glueballs can be described as long semiclassical strings, which do not probe the UV asymptotics.

On a somewhat related note, it will be interesting to clearly disentangle which of the ASA assumptions are dynamical, and which are consequences of the symmetries. As we already said in the introduction, the ASA predictions for the spin content to large extent can be thought of as the finite volume version of the Goldstone theorem. Indeed,  in section \ref{sec:spins} we used the action to derive the spin content of open ASA strings. However, this action is fixed by symmetries, so it is similar to using the chiral Lagrangian to deduce properties of pions.
On the other hand, for closed strings we used an additional step---to identify ${\cal H}_{L(R)}(N)$ with ${\cal H}_{open}(N)$, (\ref{Hlro}). This identification was motivated by the known properties of critical strings, and it will be interesting to understand whether it can be justified from general symmetry considerations, or it is a genuine dynamical property.

It is important to stress also that the two major steps required to predict the glueball quantum numbers from the ASA (the tensor square structure of section ~\ref{sec:tensor} and the 
semiclassical ansatz of section~\ref{sec:spins}) should be considered as two distinct assumptions, with a quite different status. In our view, the tensor square structure is a very robust and non-controversial  property of short strings in the absence of massive worldsheet excitations. The semiclassical ansatz is a much bolder assumption. For the sake of full disclosure, it is worth to emphasize that the analysis of lattice data presented in the paper was not fully blind, of course.
However, the tensor square structure assumption has actually been made before looking at the glueball data. On the other hand, the semiclassical ansatz emerged only after the corresponding squares were identified. Hence, at this stage the latter is more properly to be considered as an interpretation of the data, rather than an unambiguous 
theoretical prediction of the ASA.

To conclude, it may be appropriate to recall an old dream by Lord Kelvin \cite{thomson1869} who proposed that different atomic elements may be described by knotted configurations of vertex rings, and the difference in their chemical properties can be attributed to differences in the knot topology. By now we know very well that this is not the right description of atoms. However, there is a growing evidence that Kelvin's vision to large extent works as far as hadrons are concerned. Hopefully, the results reported here bring us closer towards making Kelvin's dream to come true.

\section*{Acknowledgements}
We are grateful to Andreas Athenodorou and Mike Teper for discussions and explaining the details of lattice data.
We would like to thank Raphael Flauger and Victor Gorbenko for collaboration on related topics.
We have also benefited  from discussions with Markus Luty, Sergey Monin, Sergey Sibiryakov and Pedro Vieira.
 Research at Perimeter Institute is supported by the
  Government of Canada through Industry Canada and by the Province of
  Ontario through the Ministry of Economic Development \&
  Innovation.
This work was supported in part by the NSF CAREER award PHY-1352119.

\bibliographystyle{utphys}
\bibliography{dlrrefs}
\end{document}